\documentclass[twocolumn]{autart}    % Enable this line and disable the 
\usepackage{amssymb,amsmath,amsfonts}
\usepackage{algorithm,algorithmic}
\usepackage{cite}
\usepackage{float}
\usepackage{epsfig}
\usepackage{graphicx}

\usepackage{graphics}
\usepackage{multicol}
\usepackage{lipsum}
\usepackage{subfigure}
\usepackage{url}
\usepackage{verbatim}
\usepackage{xspace}
\usepackage[usenames,dvipsnames]{color}
\usepackage{setspace}
\usepackage{dblfloatfix}

\floatstyle{ruled}
\newfloat{model}{H}{mod}
\floatname{model}{\footnotesize Model}
\newfloat{notatio}{H}{not}
\floatname{notatio}{\footnotesize Notation}

\newenvironment{varalgorithm}[1]
  {\algorithm\renewcommand{\thealgorithm}{#1}}
  {\endalgorithm}

\newenvironment{list4}{
	\begin{list}{$\bullet$}{%
			\setlength{\itemsep}{0.05cm}
			\setlength{\labelsep}{0.2cm}
			\setlength{\labelwidth}{0.3cm}
			\setlength{\parsep}{0in} 
			\setlength{\parskip}{0in}
			\setlength{\topsep}{0in} 
			\setlength{\partopsep}{0in}
			\setlength{\leftmargin}{0.16in}}}
	{\end{list}}

\newenvironment{list4a}{
	\begin{list}{$\bullet$}{%
			\setlength{\itemsep}{0.05cm}
			\setlength{\labelsep}{0.2cm}
			\setlength{\labelwidth}{0.3cm}
			\setlength{\parsep}{0in} 
			\setlength{\parskip}{0in}
			\setlength{\topsep}{0in} 
			\setlength{\partopsep}{0in}
			\setlength{\leftmargin}{0.16in}}}
	{\end{list}}

\usepackage{url,changebar,bm,xspace,dsfont}
\let\mathbb=\mathds % I much prefer the dsfont over the bbfont
\newtheorem{theorem}{Theorem}

\newtheorem{example}{\bfseries Example}
\newtheorem{remark}{Remark}

\usepackage{graphics} % for pdf, bitmapped graphics files
\usepackage{epsfig} % for postscript graphics files
\usepackage{algorithm,algorithmic}

\usepackage{amsmath,amsfonts,amssymb,amscd}
\usepackage{capt-of}

\usepackage{color}
\usepackage{cite}
\usepackage{verbatim}
\usepackage{hyperref}
\hypersetup{
     colorlinks = true,
     linkcolor = [rgb]{0,0,1},
     anchorcolor = [rgb]{0,0,1},
     citecolor = [rgb]{0.9,0.5,0},
     filecolor = [rgb]{0,0,1},
     pagecolor = [rgb]{1,1,0},
     urlcolor = [rgb]{0.9,0.5,0},
     bookmarks,
        bookmarksopen = true,
        bookmarksnumbered = true,
        breaklinks = true,
        linktocpage,
        pagebackref,
        colorlinks = true,
        linkcolor = [rgb]{0.2,0.6,0.2},
        urlcolor  = [rgb]{0,0,1},
        citecolor = [rgb]{0.9,0.5,0},
        anchorcolor = [rgb]{0.2,0.6,0.2},
        hyperindex = true,
        hyperfigures
}
\newtheorem{lemma}{\bfseries Lemma}

\begin{document}

\begin{frontmatter}

\title{Fast Quantized Average Consensus \\ over Static and Dynamic Directed Graphs\thanksref{footnoteinfo}} % Title, preferably not more 

\thanks[footnoteinfo]{An early version of the algorithm in this paper appears in the conference paper  \cite{2020:RikosHadj_IFAC}. The main differences of this paper with \cite{2020:RikosHadj_IFAC} are: (i) the proposed algorithm avoids oscillating behavior regarding the nodes' states while maintaining fast convergence speed (similar to \cite{2020:RikosHadj_IFAC}), (ii) an extended version of the proposed algorithm under a dynamically changing directed communication topology, (iii) detailed proofs for convergence for the results and avoiding oscillatory behavior are given (not provided in \cite{2020:RikosHadj_IFAC}).
Corresponding author Apostolos~I.~Rikos.}

\author[First]{Apostolos~I.~Rikos}\ead{rikos@kth.se}, 
\author[Second]{Christoforos~N.~Hadjicostis}\ead{chadjic@ucy.ac.cy}, 
\author[First]{Karl~H.~Johansson}\ead{kallej@kth.se} 

\address[First]{Division of Decision and Control Systems, KTH Royal Institute of Technology, SE-100 44 Stockholm, Sweden.}  % Please supply                                              
\address[Second]{Department of Electrical and Computer Engineering, University of Cyprus, Nicosia, Cyprus.}             % full addresses

%\begin{keyword}                          
%Quantized average consensus, event-triggered, distributed algorithms, quantization, digraphs, multi-agent systems             
%\end{keyword}                             

\begin{abstract}                          % Abstract of not more than 200 words.
In this paper we study the distributed average consensus problem in multi-agent systems with directed communication links that are subject to quantized information flow. 
%The goal of distributed average consensus is for the agents, each associated with some initial value, to obtain the average (or some value close to the average) of these initial values. 
Specifically, we present and analyze a distributed averaging algorithm which operates exclusively with quantized values (i.e., the information stored, processed and exchanged between neighboring agents is subject to deterministic uniform quantization) and relies on event-driven updates (e.g., to reduce energy consumption, communication bandwidth, network congestion, and/or processor usage). 
The main idea of the proposed algorithm is that each node (i) models its initial state as two quantized fractions which have numerators equal to the node's initial state and denominators equal to one, and (ii) transmits one fraction randomly while it keeps the other stored. 
Then, every time it receives one or more fractions, it averages their numerators with the numerator of the fraction it stored, and then transmits them to randomly selected out-neighbors.
We characterize the properties of the proposed distributed algorithm and show that its execution, on any static and strongly connected digraph, allows each agent to reach in finite time a fixed state that is equal (within one quantisation level) to the average of the initial states. 
We extend the operation of the algorithm to achieve finite-time convergence in the presence of a dynamic directed communication topology subject to some connectivity conditions. 
Finally, we provide examples to illustrate the operation, performance, and potential advantages of the proposed algorithm. 
We compare against state-of-the-art quantized average consensus algorithms and show that our algorithm's convergence speed significantly outperforms most existing protocols. 
\end{abstract}


\end{frontmatter}

% ===============================================
%
%
% INTRODUCTION
%
%
% ===============================================
\section{Introduction}\label{intro}

In recent years, there has been a growing interest for control and coordination of networks consisting of multiple agents, like groups of sensors \cite{2005:XiaoBoydLall} or mobile autonomous agents \cite{2004:Murray}. 
A problem of particular interest in distributed control is the \textit{consensus} problem where the objective is to develop distributed algorithms that can be used by a group of agents in order to reach agreement to a common decision. 
The agents start with different initial states/information and are allowed to communicate locally via inter-agent information exchange under some constraints on connectivity. 
Consensus processes play an important role in many problems, such as leader election \cite{1996:Lynch},  motion coordination \cite{2005:Olshevsky_Tsitsiklis, 2004:Murray}, and clock synchronization \cite{2007:Gamba}.

%The consensus problem also arises in a number of applications including coordination of UAVs (e.g., aligning the agents’ directions of motion), information processing in sensor networks, and distributed optimization (e.g., agreeing on the estimates of some unknown parameters). 
%The averaging problem is a special case in which the goal is to compute the exact average of the initial values of the agents.

One special case of the consensus problem is the distributed averaging problem, where each agent (initially endowed with a numerical state) can send/receive information to/from other agents in its neighborhood and update its state iteratively, so that eventually, all agents compute the average of the initial states. 
Average consensus is an important problem and has been studied extensively, primarily in settings where each agent processes and transmits real-valued states with infinite precision \cite{2018:BOOK_Hadj, 2005:Olshevsky_Tsitsiklis, 2008:Sundaram_Hadjicostis, 2013:Themis_Hadj_Johansson, 2011:Morse_Yu}. 
However, most existing average consensus algorithms are only able to guarantee asymptotic convergence, implying that they cannot be readily applied to real-world distributed control and coordination applications.
Furthermore, constraints on the bandwidth of communication links and the capacity of physical memories require both communication and computation to be performed assuming finite precision.
For these reasons, researchers have studied the case where network links can only allow messages of limited length to be transmitted between agents, effectively extending techniques for average consensus towards the direction of quantized average consensus \cite{2007:Aysal_Rabbat, 2012:Lavaei_Murray, 2007:Basar, 2008:Carli_Zampieri, 2013:GarciaCasbeer, 2016:Chamie_Basar, 2011:Cai_Ishii}. 
In addition, the demand for more efficient usage of network resources, has lead to an increasing interest for novel event-triggered algorithms for distributed control  \cite{2013:Dimarogonas_Johansson, 2016:nowzari_cortes, 2012:Liu_Chen}.

Distributed algorithms that achieve quantized average consensus in an event-driven fashion have a wide variety of applications. 
They can be used as the basis for various encoding schemes, such as quantized privacy protocols for guaranteeing additional levels of security without significantly increasing communication overhead \cite{2020:RikosThemisJohHadj_CDC, 2019:Ruan_Wang}. 
Furthermore, in recent years there is a tremendous growth in distributed optimization \cite{2005:Rabbat_Nowak, 2008:Nedic_Tsitsiklis, 2020:Khatana_Salapaka, 2020:Themis_Kalyvianaki} and machine learning algorithms \cite{2018:Jiang_Agrawal, 2019:Giannakis_Yan}. 
The distributed operation of these algorithms over directed graphs requires exchange of the agents’
states without any error in order to guarantee convergence to a desired solution.
However, as the network size becomes larger, e.g., to speed up the training of deep learning algorithms, the communication overhead of each iteration becomes a major bottleneck. 
Quantization and event-driven communication are effective approaches to tackle this issue, since reduction of communication and processing costs leads to bandwidth and energy efficient algorithms. 
The emerging importance of the aforementioned approaches can be further seen in various recent works \cite{2020:Shlezinger, 2020:Sun_Giannakis, 2020:Elgabli_Aggarwal, 2020:Jadbabaie} where researchers present machine learning algorithms, which employ quantization strategies to tackle the large communication overhead and reduce communication payload size, while maintaining fast convergence rates and possible privacy preserving guarantees. 
%As a result, due to the particular importance of communication efficiency and convergence speed, as well as the need for providing encoding and privacy guarantees, there is an increasing demand for fast distributed coordination algorithms to process and transmit quantized parameters, while serving as backbones for privacy, optimization and machine learning. \todo{REMOVE LAST PHRASE?}

\subsection{Literature Review}\label{review}

%In the remainder of this section, we review the existing literature on algorithms for distributed averaging under quantized communication, classifying them according to whether they converge in a probabilistic or a deterministic fashion. 

In recent years, quite a few {\em probabilistic} distributed algorithms for averaging under quantized communication, have been proposed. 
Specifically, the probabilistic quantizer in \cite{2007:Aysal_Rabbat} converges to a common state with a random quantization level for the case where the topology forms a directed graph. 
In \cite{2010:Kar_Moura} the authors present a distributed algorithm which adds a dither to the agents' measurements (before the quantization process) and they show that the mean square error can be made arbitrarily small. 
In \cite{2011:Benezit_Vetterli} the authors present a distributed algorithm which guarantees that all agents reach consensus to a value on the interval in which the average lies after a finite number of time steps. 
In \cite{2012:Lavaei_Murray} the authors present a quantized gossip algorithm which deals with the distributed averaging problem over a connected weighted graph, and calculates lower and upper bounds on the expected value of the convergence time, which depend on the principal submatrices of the Laplacian matrix of the weighted graph.

The available literature concerning {\em deterministic} distributed algorithms for averaging under quantized communication comprises less publications. 
In \cite{2011:Li_Zhang}, the authors present a distributed averaging algorithm with dynamic encoding and
decoding schemes. 
They show that for a connected undirected dynamic graph, average consensus is achieved asymptotically with
as few as one bit of information exchange between each pair of adjacent agents at each time step, and the 
convergence rate is asymptotic and depends on the number of network nodes, the number of quantization levels and the synchronizability of the network.
In \cite{2013:Thanou_Frossard} the authors present a novel quantization scheme for solving the average consensus problem when sensors exchange quantized state information. 
The proposed scheme is based on progressive reduction of the range of a uniform quantizer and leads to progressive refinement of the information exchanged by the sensors.
In \cite{2008:Carli_Zampieri} the authors derive bounds on the rate of convergence to average consensus for a team of mobile agents exchanging information over time-invariant or randomly time-varying communication networks with symmetries. 
Furthermore, they study the control performance when agents also exchange logarithmically quantized data over static communication topologies with symmetries.
In \cite{2009:Nedic} the authors study distributed algorithms for the averaging problem over networks with dynamic topologies, with a focus on tight bounds on the convergence time of a general class of averaging algorithms.
They consider algorithms for the case where agents can exchange and store continuous or quantized states, establish a tight convergence rate, and show that these algorithms guarantee convergence within some error from the average of the initial states; this error depends on the number of quantization levels.

Recent papers have studied the quantized average consensus problem with the additional constraint that the state of each node is an integer value. 
In \cite{2007:Basar} the authors present a probabilistic algorithm which allows every agent to reach quantized consensus almost surely over a static and undirected communication topology, while in \cite{2016:Etesami_Basar} and \cite{2016:Basar_Olshevsky} the authors analyze and further improve its convergence rate. 
In \cite{2011:Cai_Ishii} a probabilistic algorithm was proposed to solve the quantized consensus problem for static directed graphs for the case where the agents exchange quantized information and store the changes of their states in an additional (also quantized) variable called ``surplus''. 
The authors of \cite{2016:Chamie_Basar} present a deterministic distributed averaging algorithm subject to quantization on the links and show that, depending on initial conditions, the system either converges in finite time to quantized consensus, or the nodes enter into a periodic behaviour with their states oscillating around the average.
In \cite{2017:MouCasbeer}, the authors present two distributed algorithms, one for fixed tree graphs with finite time convergence and one for dynamic directed graphs with exponential convergence. 
The algorithms proposed in this work calculate the initial average as the the ratio of two scaled sums obtained by running in parallel two iterations. 
In \cite{2018:RikosHadj_CDC, 2020:Rikos_Quant_Cons} the authors present two distributed algorithms, one probabilistic and one deterministic, which calculate the exact quantized average of the initial states (i.e., there is no quantization error) in a finite number of time steps, which is explicitly calculated. 
In \cite{2020:RikosHadj_IFAC} the authors present a distributed randomized algorithm which calculates the quantized average of the initial states with high probability. 
The algorithm is shown to outperform other algorithms but the states of the nodes exhibit oscillating behavior (between the ceiling and the floor of the real-valued average of the initial states).

\subsection{Main Contributions}

In this paper, we present a novel distributed average consensus algorithm in which processing, storing, and exchange of information between neighboring agents is event-driven and  subject to uniform quantization. 
The main idea behind the proposed algorithm is that initially each node stores two fractions. 
Each fraction has numerator equal to the node's initial quantized state and denominator equal to one. 
Then, the node transmits one fraction randomly while it keeps the other stored. 
Every time it receives one or more fractions, it averages their numerators with the numerator of the fraction it keeps stored, and then transmits them to randomly selected out-neighbors. 
Note here that most work dealing with quantization has concentrated on the scenario where the agents have real-valued states but can only transmit quantized values through limited rate channels (e.g., \cite{2008:Carli_Zampieri, 2016:Chamie_Basar}). 
By contrast, our setup covers the case where the states are stored in digital memories of finite capacity (as in \cite{2009:Nedic, 2007:Basar, 2011:Cai_Ishii}). 
Specifically, we assume that states are integer-valued (which comprises a class of quantization effects such as uniform quantization) and the control actuation of each node is event-based, which enables more efficient use of available resources.

The main contribution of this paper is fourfold. 
\\ A. We introduce a novel distributed algorithm that allows all agents to reach quantized average consensus in finite time with high probability (Algorithm~\ref{algorithm_prob_no_oscil}).
\\ B. We show that, unlike existing algorithms in the literature, this algorithm allows every agent to calculate either the ceiling or the floor of the real average of the initial states with no oscillations in finite time (Theorem~\ref{Theorem2_prob_no_oscill}). 
\\ C. We present experimental results in which we compare the proposed algorithm against existing schemes and observe that its convergence speed significantly outperforms most finite-time distributed algorithms for average consensus under quantized communication. 
\\ D. We present an enhanced version of the algorithm (Algorithm~\ref{algorithm_prob_no_oscil_dynamic}) which achieves fast oscillation-free convergence in finite time in the presence of a dynamically changing communication topology (Theorem~\ref{Theorem2_prob_no_oscill_dynamic}).

\subsection{Outline}

The remainder of this paper is organized as follows. 
In Section~\ref{notation}, we introduce the notation used throughout the paper, while in Section~\ref{probForm} we formulate the quantized average consensus problem.
In Section~\ref{algorithm_no_oscil}, we present a probabilistic distributed algorithm which operates over static digraphs and allows the agents to reach consensus to the quantized average of the initial states, in finite time, with probability one. 
We demonstrate its performance with an illustrative example, and we analyze its operation and establish its finite time termination. 
In Section~\ref{algorithm_no_oscil_dyn_digr} we present an enhanced version of our algorithm which allows every agent to reach consensus to the quantized average of the initial states in the presence of a dynamic communication topology. 
In Section~\ref{results}, we present simulation results and comparisons against the current state-of-the-art. 
We conclude in Section~\ref{future} with a brief summary and remarks about future work.

% ===============================================
%
%
% NOTATION
%
%
% ===============================================
\section{Preliminaries}\label{notation}

\subsection{Notation}\label{graph_notation}

The sets of real, rational, integer and natural numbers are denoted by $ \mathbb{R}, \mathbb{Q}, \mathbb{Z}$ and $\mathbb{N}$, respectively. 
The symbol $\mathbb{Z}_+$ denotes the set of nonnegative integers and the symbol $\mathbb{N}_0$ denotes the set of natural numbers that also includes zero. 
For any real number $a \in \mathbb{R}$, the floor $\lfloor a \rfloor$ denotes the greatest integer less than or equal to $a$ while the ceiling $\lceil a \rceil$ denotes the least integer greater than or equal to $a$. 

%Consider a network of $n$ ($n \geq 2$) agents communicating only with their immediate neighbors. 
The communication topology is a network of $n$ ($n \geq 2$) agents communicating only with their immediate neighbors, and can be captured by a directed graph (digraph), called \textit{communication digraph}, which is either \textit{static} (i.e., it does not change over time) or \textit{dynamic} (i.e., it changes over time). 

A \textit{static} digraph is defined as $\mathcal{G}_d = (\mathcal{V}, \mathcal{E})$, where $\mathcal{V} =  \{v_1, v_2, \dots, v_n\}$ is the set of nodes (representing the agents of the multi-agent system) and $\mathcal{E} \subseteq \mathcal{V} \times \mathcal{V} - \{ (v_j, v_j) \ | \ v_j \in \mathcal{V} \}$ is the set of edges (self-edges excluded). 
A directed edge from node $v_i$ to node $v_j$ is denoted by $m_{ji} \triangleq (v_j, v_i) \in \mathcal{E}$, and captures the fact that node $v_j$ can receive information from node $v_i$ (but not the other way around). 
We adopt the common assumption that the given static digraph $\mathcal{G}_d = (\mathcal{V}, \mathcal{E})$ is \textit{strongly connected} (i.e., for each pair of nodes $v_j, v_i \in \mathcal{V}$, $v_j \neq v_i$, there exists a directed \textit{path} from $v_i$ to $v_j$). 
The subset of nodes that can directly transmit information to node $v_j$ is called the set of in-neighbors of $v_j$ and is represented by $\mathcal{N}_j^- = \{ v_i \in \mathcal{V} \; | \; (v_j,v_i)\in \mathcal{E}\}$, while the subset of nodes that can directly receive information from node $v_j$ is called the set of out-neighbors of $v_j$ and is represented by $\mathcal{N}_j^+ = \{ v_l \in \mathcal{V} \; | \; (v_l,v_j)\in \mathcal{E}\}$. 
The cardinality of $\mathcal{N}_j^-$ is called the \textit{in-degree} of $v_j$ and is denoted by $\mathcal{D}_j^-$ (i.e., $\mathcal{D}_j^- = | \mathcal{N}_j^- |$), while the cardinality of $\mathcal{N}_j^+$ is called the \textit{out-degree} of $v_j$ and is denoted by $\mathcal{D}_j^+$ (i.e., $\mathcal{D}_j^+ = | \mathcal{N}_j^+ |$). 

In dynamic digraphs we assume that the set of nodes is fixed while the set of edges among them might change at various points in time. 
Specifically, a \textit{dynamic} digraph is defined as $\mathcal{G}_d[k] = (\mathcal{V}, \mathcal{E}[k])$, where $\mathcal{V} =  \{v_1, v_2, \dots, v_n\}$ is the set of nodes (representing the agents of the multi-agent system) and $\mathcal{E}[k] \subseteq \mathcal{V} \times \mathcal{V} - \{ (v_j, v_j) \ | \ v_j \in \mathcal{V} \}$ is the set of edges at time step $k$ (self-edges excluded). 
A directed edge from node $v_i$ to node $v_j$ is denoted by $m_{ji} \triangleq (v_j, v_i) \in \mathcal{E}[k]$, and captures the fact that node $v_j$ can receive information from node $v_i$ (but not the other way around) at time step $k$. 
This means that at each time instant $k$, each node $v_j$ has possibly different sets of in- and out-neighbors, denoted respectively by $\mathcal{N}_j^-[k]$ and $\mathcal{N}_j^+[k]$ and defined as $\mathcal{N}_j^-[k] = \{ v_i \in \mathcal{V} \; | \; (v_j,v_i) \in \mathcal{E}[k]\}$ and $\mathcal{N}_j^+[k] = \{ v_l \in \mathcal{V} \; | \; (v_l,v_j) \in \mathcal{E}[k]\}$. 
Furthermore, the cardinality of $\mathcal{N}_j^-[k]$, at time step $k$, is called the \textit{in-degree} of $v_j$ and is denoted by $\mathcal{D}_j^-[k] = | \mathcal{N}_j^-[k] |$, while the cardinality of $\mathcal{N}_j^+[k]$, at time step $k$, is called the \textit{out-degree} of $v_j$ and is denoted by $\mathcal{D}_j^+[k] = | \mathcal{N}_j^+[k] |$. 
Given a collection of digraphs $\mathcal{G}_d[k] = (\mathcal{V}, \mathcal{E}[k])$ for $k = 1, 2, ..., m$, where $m \in \mathbb{N}$, the \textit{union graph} is defined as $\mathcal{G}^{1, 2, ..., m}_d = (\mathcal{V}, \cup_{k = 1}^{m} \mathcal{E}[k])$. 
A collection of digraphs is said to be \textit{jointly strongly connected}, if its corresponding union graph $\mathcal{G}^{1, 2, ..., m}_d$ forms a strongly connected digraph.
%Note here that we have a strongly connected digraph if (i) at least one of the digraphs in the collection is strongly connected or (ii) one of the digraphs forming the union is strongly connected. 
%This means that in the first case we need at least one $\mathcal{G}_d[k_0]$ for $k_0 \in [1, m]$ to be strongly connected, while in the second case we need $\mathcal{G}^{k_1, ..., k_1'}_d = (\mathcal{V}, \cup_{k = k_1}^{k = k_1'} \mathcal{E}[k])$, where $k_1 < k_1'$ and $k_1, k_1' \in [1, m]$, to be strongly connected.

\subsection{Agent Operation}

With respect to quantization of information flow each node $v_j \in \mathcal{V}$ maintains, at time step $k$, $5 + 2 \mathcal{D}_j^+$ variables, as follows: 
\\ \noindent
(i) The mass variables $y_j[k], z_j[k]$, where $y_j[k] \in \mathbb{Z}$ and $z_j[k] \in \mathbb{N}_0$, are used for processing and calculating the average of the initial states.
\\ \noindent
(ii) The state variables $y^s_j[k], z^s_j[k], q_j^s[k]$, where $y^s_j[k] \in \mathbb{Z}$, $z^s_j[k] \in \mathbb{N}$ and $q_j^s[k] \in \mathbb{Z}$ (with $q_j^s[k] = \lfloor \frac{y_j^s[k]}{z_j^s[k]} \rfloor$ or $q_j^s[k] = \lceil \frac{y_j^s[k]}{z_j^s[k]} \rceil$), are used for storing the values of the received mass variables and for calculating the state variable $q_j^s$, which is the variable that becomes equal to the quantized average of the initial states.
\\ \noindent
(iii) The transmission variables $c^{y}_{lj}[k]$ and $c^{z}_{lj}[k]$ for each $v_l \in \mathcal{N}^+_j[k]$, where $c^{y}_{lj}[k] \in \mathbb{Z}$ and $c^{z}_{lj}[k] \in \mathbb{N}_0$, are used for transmitting $v_j$'s mass variables towards its out-neighbors.

%The aggregate states of the mass and state variables are denoted by $y[k] = [y_1[k] \ ... \ y_n[k]]^{\rm T} \in \mathbb{Z}^n$, $z[k] = [z_1[k] \ ... \ z_n[k]]^{\rm T} \in \mathbb{N}_0^n$, $y^s[k] = [y^s_1[k] \ ... \ y^s_n[k]]^{\rm T} \in \mathbb{Z}^n$, $z^s[k] = [z^s_1[k] \ ... \ z^s_n[k]]^{\rm T} \in \mathbb{N}^n$, and $q^s[k] = [q^s_1[k] \ ... \ q^s_n[k]]^{\rm T} \in \mathbb{Z}^n$ respectively.  \todo{is this paragraph useful?}

%\begin{remark} \todo{is this remark useful?}
%Note that during its operation, each node processes and transmits integer values via available communication links, while, following \cite{2007:Basar, 2011:Cai_Ishii}, we assume that the state variables maintained at each node are also integer valued. 
%This abstraction subsumes a class of quantization effects (e.g., uniform quantization).
%\end{remark}

\subsection{Transmission Strategy}\label{trans_str}

In the static communication topology case, in order to randomly determine which out-neighbor to transmit to, each node $v_j$ assigns a nonzero probability $b_{lj}$ to each of its outgoing edges $m_{lj}$ (including a virtual self-edge), where $v_l \in \mathcal{N}^+_j \cup \{ v_j \}$. 
This probability assignment for all nodes can be captured by an $n \times n$ column stochastic matrix $\mathcal{B} = [b_{lj}]$. 
A simple choice would be to set these probabilities to be equal, i.e.,
\begin{align*}
b_{lj} = \left\{ \begin{array}{ll}
         \frac{1}{1 + \mathcal{D}_j^+}, & \mbox{if $v_{l} \in \mathcal{N}_j^+ \cup \{v_j\}$,}\\
         0, & \mbox{otherwise.}\end{array} \right. 
\end{align*}
Each nonzero entry $b_{lj}$ of matrix $\mathcal{B}$ represents the probability of node $v_j$ transmitting towards out-neighbor $v_l \in \mathcal{N}^+_j$ through the edge $m_{lj}$, or transmitting to itself (i.e., performing no transmission with probability $b_{jj}$). 
Note here that, for every node $v_j$, the probability $b_{jj}$ being positive implies that the matrix $\mathcal{B}$ becomes primitive (recall that the underlying static digraph is assumed to be strongly connected), which is of particular importance for the results presented in this paper. 

In the dynamic communication topology case, we still have that each node randomly decides which out-neighbor to transmit to. 
Specifically, each node $v_j$ assigns a nonzero probability $b_{lj}[k]$ to each of its outgoing edges $m_{lj}[k]$ (including a virtual self-edge) at each time step $k$, where $v_l \in \mathcal{N}^+_j[k] \cup \{ v_j \}$. 
This probability assignment for all nodes can be captured, at each time step $k$, by an $n \times n$ column stochastic matrix $\mathcal{B}[k] = [b_{lj}[k]]$. 
Again, a simple choice would be to set these probabilities to be equal, i.e.,
\begin{align*}
b_{lj}[k] = \left\{ \begin{array}{ll}
         \frac{1}{1 + \mathcal{D}_j^+[k]}, & \mbox{if $v_{l} \in \mathcal{N}_j^+[k] \cup \{v_j\}$,}\\
         0, & \mbox{otherwise.}\end{array} \right. 
\end{align*}
Each nonzero entry $b_{lj}[k]$ of matrix $\mathcal{B}[k]$ represents the probability of node $v_j$ transmitting towards out-neighbor $v_l \in \mathcal{N}^+_j[k]$ through the edge $m_{lj}[k]$ at time step $k$, or transmitting to itself (i.e., performing no transmission with probability $b_{jj}[k]$).
Let us note here that the dynamic nature of the underlying communication topology implies that the matrix $\mathcal{B}[k]$ is not necessarily primitive at each time step $k$ (whereas for a static strongly connected topology, the corresponding $\mathcal{B}$ will necessarily be primitive). 
%However, since a collection of dynamic digraphs is said to be jointly strongly connected, i.e., the corresponding union graph $\mathcal{G}^{1, 2, ..., m}_d$ forms a strongly connected digraph (see Section~\ref{graph_notation}), then we have that the matrix $\mathcal{B}^{1, 2, ..., m}$ (that captures probability assignment for all nodes for a collection of dynamic digraphs) becomes primitive. 
%\todo{is the last phrase useful?}

%\textbf{Transmission Delays:}
%We assume that a transmission from node $v_j$ to node $v_l$ at time step $k$ undergoes an \textit{a priori unknown} delay $\tau^{(j)}_{lj}[k]$. 
%More specifically we assume that $\tau^{(j)}_{lj}[k]$ is an integer that satisfies $0 \leq \tau^{(j)}_{lj}[k] \leq \overline{\tau}_{lj} < \infty$ where the maximum delay is denoted by $\overline{\tau} = \max_{(v_l,v_j) \in \mathcal{E}}\overline{\tau}_{lj}$.  
%Under the above delay model, the packet sent by node $v_j$ at time step $k$, is received by node $v_l$ at time step $k + \tau^{(j)}_{lj}[k]$. 

% ===============================================
%
%
% PROBLEM
%
%
% ===============================================
\section{Problem Formulation}\label{probForm}

Consider a digraph $\mathcal{G}_d = (\mathcal{V}, \mathcal{E})$, where each node $v_j \in \mathcal{V}$ has an initial quantized state $y_j[0]$ (for simplicity, we take $y_j[0] \in \mathbb{Z}$) and $q$ is the real average of the initial states:  
\begin{equation}\label{real_av}
q = \frac{\sum_{l=1}^{n}{y_l[0]}}{n} .
\end{equation}
In this paper, we aim to develop distributed algorithms that address the following problems \textbf{P1} -- \textbf{P2}.

\noindent
\textbf{P1.}
When the digraph $\mathcal{G}_d$ is static and strongly connected, the algorithm allows the nodes to obtain, after a finite number of steps, a quantized state $q^s$ which is equal to the ceiling or the floor of the actual average $q$ of the initial states in \eqref{real_av}. 
Specifically, we require that there exists $k_0$ so that for every $v_j \in \mathcal{V}$ we have
\begin{equation}\label{alpha_q_no_oscill}
( q^s_j[k] = \lfloor q \rfloor \ \ \text{for} \ \ k \geq k_0 ) \ \ \text{or} \ \ ( q^s_j[k] = \lceil q \rceil \ \ \text{for} \ \ k \geq k_0).
\end{equation}

\noindent
\textbf{P2.}
When the digraph $\mathcal{G}_d[k]$ is dynamic and jointly strongly connected, we require that there exists $k_0$ so that \eqref{alpha_q_no_oscill} holds.

The quantized average $q^s$ is defined as the ceiling $\lceil q \rceil$ or the floor $\lfloor q \rfloor$ of the true average $q$ of the initial states in \eqref{real_av}. 
Let $S \triangleq \mathbf{1}^{\rm T} y[0]$, where $\mathbf{1} = [1 \ ... \ 1]^{\rm T}$ is the vector of all ones, and let $y[0] = [y_1[0] \ ... \ y_n[0]]^{\rm T}$ be the vector of the quantized initial states. 
We can write $S$ uniquely as 
\begin{equation}\label{equation_mod}
S = nL + R
\end{equation}
where $L$ and $R$ are both integers and $0 \leq R < n$. 
Thus, we have that either $L$ or $L+1$ may be viewed as an integer approximation of the average of the initial states $q = S/n$ (which may not be integer in general).

\begin{remark}
Note here that our definition of quantized average consensus is different than in some literature \cite{2007:Basar, 2011:Cai_Ishii, 2020:RikosHadj_IFAC, 2020:Rikos_Quant_Cons, 2016:Chamie_Basar}.
We require that all agents states converge to a specific integer (either $\lfloor q \rfloor$ or $\lceil q \rceil$ where $q$ satisfies \eqref{real_av}). 
Apart from \cite{2011:Cai_Ishii}, this cannot be achieved in existing finite-time algorithms since they either exhibit oscillating behavior of the agent states between the values $\lfloor q \rfloor$ or $\lceil q \rceil$ \cite{2007:Basar, 2020:RikosHadj_IFAC, 2016:Chamie_Basar}, or calculate the average in the  form of a quantized fraction \cite{2020:Rikos_Quant_Cons}. 
\end{remark}

\section{Quantized Averaging over Static Digraphs}\label{algorithm_no_oscil}

%In this section we propose a probabilistic distributed information exchange protocol which addresses problem \textbf{(P1)} presented in Section~\ref{probForm}. 
%%the nodes transmit and receive quantized messages so that they reach quantized average consensus on their initial values after a finite number of steps. 
%This probabilistic quantized mass transfer process is detailed as Algorithm~\ref{algorithm_prob} below (for the case where $b_{lj} = 1/(1+\mathcal{D}_j^+)$ for $v_l \in \mathcal{N}_j^+ \cup \{ v_j \}$ and $b_{lj}=0$ otherwise). 

In this section we propose a probabilistic distributed information exchange algorithm which addresses problem \textbf{(P1)} presented in Section~\ref{probForm}. 
This is detailed as Algorithm~\ref{algorithm_prob_no_oscil} below (for the case where $b_{lj} = 1/(1+\mathcal{D}_j^+)$ for $v_l \in \mathcal{N}_j^+ \cup \{ v_j \}$ and $b_{lj}=0$ otherwise).

\begin{varalgorithm}{1}
\caption{Quantized Average Consensus over Static Digraphs}
\noindent \textbf{Input} A strongly connected digraph $\mathcal{G}_d = (\mathcal{V}, \mathcal{E})$ with $n=|\mathcal{V}|$ nodes and $m=|\mathcal{E}|$ edges. 
Each node $v_j\in \mathcal{V}$ has an initial state $y_j[0] \in \mathbb{Z}$. \\
\textbf{Initialization:} Each node $v_j \in \mathcal{V}_p$ does the following:
\begin{list4}
\item[$1)$] Assigns a nonzero probability $b_{lj}$ to each of its outgoing edges $m_{lj}$, where $v_l \in \mathcal{N}^+_j \cup \{v_j\}$, as follows  
\begin{align*}
b_{lj} = \left\{ \begin{array}{ll}
         \frac{1}{1 + \mathcal{D}_j^+}, & \mbox{if $l = j$ or $v_{l} \in \mathcal{N}_j^+$,}\\
         0, & \mbox{if $l \neq j$ and $v_{l} \notin \mathcal{N}_j^+$.}\end{array} \right. 
\end{align*}
\item[$2)$] Sets $y_j[0] := 2y_j[0]$, $z_j[0] = 2$. 
\end{list4}
\textbf{Iteration:} For $k=0,1,2,\dots$, each node $v_j \in \mathcal{V}_p$, does the following:
\begin{list4}
\item[$1)$] \textbf{if} $z_j[k] > 1$, \textbf{then} 
\begin{list4a}
\item[$1.1)$] sets $z^s_j[k] = z_j[k]$, $y^s_j[k] = y_j[k]$, 
$
q^s_j[k] = \Bigl \lfloor \frac{y^s_j[k]}{z^s_j[k]} \Bigr \rfloor \ ;
$
\item[$1.2)$] sets (i) $mas^y[k] = y_j[k]$, $mas^z[k] = z_j[k]$; (ii) $c^{y}_{lj}[k] = 0$, $c^{z}_{lj}[k] = 0$, for every $v_l \in \mathcal{N}^+_j \cup \{v_j\}$; (iii) $\delta = \lfloor mas^y[k] / mas^z[k] \rfloor$, $mas^{rem}[k]= y_j[k] - \delta \ mas^z[k]$; 
\item[$1.3)$] \textbf{while} $mas^z[k] > 1$, \textbf{then}
\begin{list4a}
\item[$1.3a)$] chooses $v_l \in \mathcal{N}^+_j \cup \{v_j\}$ randomly according to $b_{lj}$; 
\item[$1.3b)$] sets (i) $c^{z}_{lj}[k] := c^{z}_{lj}[k] + 1$, $c^{y}_{lj}[k] := c^{y}_{lj}[k] + \delta$; (ii) $mas^z[k] := mas^z[k] - 1$, $mas^y[k] := mas^y[k] - \delta$;
\item[$1.3c)$] \textbf{if} $mas^{rem}[k] > 0$, \textbf{then} sets $c^{y}_{lj}[k] := c^{y}_{lj}[k] + 1$, $mas^{rem}[k] := mas^{rem}[k]- 1$. 
\end{list4a}
\item[$1.4)$] sets $c^{y}_{jj}[k] := c^{y}_{jj}[k] + mas^y[k]$, $c^{z}_{jj}[k] := c^{z}_{jj}[k] + mas^z[k]$; 
\item[$1.5)$] \textbf{if} $c^{z}_{lj}[k] > 0$ \textbf{then}, transmits $c^{y}_{lj}[k]$, $c^{z}_{lj}[k]$ to out-neighbor $v_l$, for every $v_l \in \mathcal{N}^+_j$.
\end{list4a}
\item \textbf{else if} $z_j[k] \leq 1$, \textbf{then} sets $c^{y}_{jj}[k] = y[k]$, $c^{z}_{jj}[k] = z[k]$; 
\item[$2)$] receives $c^{y}_{ji}[k]$, $c^{z}_{ji}[k]$ from $v_i \in \mathcal{N}_j^-$ and sets 
\begin{equation}\label{no_del_eq_y_no_oscil}
y_j[k+1] = c^{y}_{jj}[k] + \sum_{v_i \in \mathcal{N}_j^-}  w_{ji}[k] \ c^{y}_{ji}[k] ,
\end{equation} 
\begin{equation}\label{no_del_eq_z_no_oscil}
z_j[k+1] = c^{z}_{jj}[k] + \sum_{v_i \in \mathcal{N}_j^-} w_{ji}[k] \ c^{z}_{ji}[k] ,
\end{equation}
where $w_{ji}[k] = 1$ if node $v_j$ receives $c^{y}_{ji}[k]$, $c^{z}_{ji}[k]$ from $v_i \in \mathcal{N}_j^-$ at iteration $k$ (otherwise $w_{ji}[k] = 0$). 
\end{list4}
\textbf{Output:} \eqref{alpha_q_no_oscill} holds for every $v_j \in \mathcal{V}$. 
\label{algorithm_prob_no_oscil}
\end{varalgorithm}

The intuition behind Algorithm~\ref{algorithm_prob_no_oscil} is  the following. 
Initially, each node $v_j$ doubles its mass variables (i.e., it sets $y_j[0] := 2 y_j[0]$ and $z_j[0] = 2$). 
Note that this change has no effect on the average calculation (since both the sums of the initial $y$-values and  $z$-values are doubled so that $(\sum_{j=1}^{n}{y_j[0]}) / (\sum_{j=1}^{n}{z_j[0]})$ remains unchanged and equal to $q$ in \eqref{real_av}).
Then, at each time step $k$, each node $v_j$ checks its mass variable $z_j[k]$ and if $z_j[k] > 1$ (i) it updates its state variables to be equal to the mass variables and (ii) it \textit{splits} $y_j[k]$ into $z_j[k]$ equal integer pieces (with the exception of some pieces whose value might be greater than others by one). 
It chooses one piece with minimum $y$-value and transmits it to itself, and it transmits each of the remaining $z_j[k]-1$ pieces to randomly selected out-neighbors or to itself. 
Finally, it receives the values $y_i[k]$ and $z_i[k]$ from its in-neighbors, sums them with its stored $y_j[k]$ and $z_j[k]$ values and repeats the operation.

We next provide an example to illustrate the operation of Algorithm~\ref{algorithm_prob_no_oscil}.

\begin{example}\label{Ex2}

Consider the strongly connected digraph $\mathcal{G}_d = (\mathcal{V}, \mathcal{E})$ in Fig.~\ref{prob_example} (borrowed from \cite{2018:RikosHadj_CDC}), with $\mathcal{V} = \{ v_1, v_2, v_3, v_4 \}$ and $\mathcal{E} = \{ m_{21}, m_{31}, m_{42}, m_{13}, m_{23}, m_{34} \}$, where nodes have initial quantized states $y_1[0] = 5$, $y_2[0] = 3$, $y_3[0] = 7$, and $y_4[0] = 2$, respectively. 
The actual average $q$ of the initial states of the nodes, is equal to $q = 4.25$ which means that the quantized state $q^s$ is equal to $q^s = 4$ or $q^s = 5$ (see Section~\ref{probForm}).

\begin{figure}[t]
\begin{center}
\includegraphics[width=0.28\columnwidth]{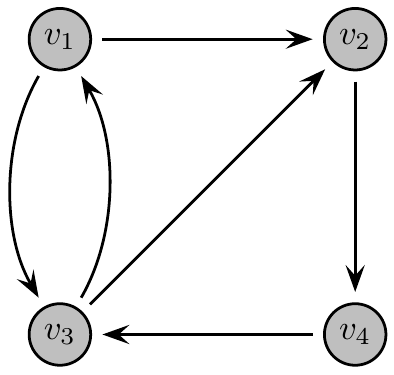}
\caption{Example of digraph for quantized averaging.}
\label{prob_example}
\end{center}
\end{figure}

Each node $v_j \in \mathcal{V}$ follows the Initialization steps ($1-2$) in Algorithm~\ref{algorithm_prob_no_oscil}, assigning to each of its outgoing edges $v_l \in \mathcal{N}_j^+ \cup \{v_j\}$ a nonzero probability value $b_{lj}$ equal to $b_{lj} = \frac{1}{1 + \mathcal{D}_j^+}$. 
The assigned values can be seen in the following matrix
\[
\mathcal{B}
=
\begin{bmatrix}
    \frac{1}{3} & \hspace{.3cm} 0 & \hspace{.3cm} \frac{1}{3} & \hspace{.3cm} 0\\ \vspace{-0.35cm} \\
    \frac{1}{3} & \hspace{.3cm} \frac{1}{2} & \hspace{.3cm} \frac{1}{3} & \hspace{.3cm} 0\\ \vspace{-0.35cm} \\
    \frac{1}{3} & \hspace{.3cm} 0 & \hspace{.3cm} \frac{1}{3} & \hspace{.3cm} \frac{1}{2}\\ \vspace{-0.35cm} \\
    0 & \hspace{.3cm} \frac{1}{2} & \hspace{.3cm} 0 & \hspace{.3cm} \frac{1}{2}\\ \vspace{-0.35cm} \\
\end{bmatrix}
.
\] 
Furthermore, nodes $v_1$, $v_2$, $v_3$, $v_4$, set $y_1[0] = 10$, $z_1[0] = 2$, $y_2[0] = 6$, $z_2[0] = 2$, $y_3[0] = 14$, $z_3[0] = 2$, $y_4[0] = 4$, $z_4[0] = 2$, respectively. 

For the execution of Algorithm~\ref{algorithm_prob_no_oscil}, at time step $k=0$, each node $v_j$ calculates its state variables $y^s_j[0]$, $z^s_j[0]$ and $q^s_j[0]$ as shown in Table~\ref{tableProb_no_oscill}. 
Then, every node $v_j$ calculates its transmission variables $c^{y}_{lj}[k]$ and $c^{z}_{lj}[k]$ for every $v_l \in \mathcal{N}_j^+ \cup \{v_j\}$. 
Specifically, each node $v_j$ splits $y_j[0]$ in $2$ equal pieces (since $z_j[0]=2$), and keeps one piece (which has the minimum value) for itself and transmits the other piece to a randomly chosen out-neighbor or itself according to the matrix $\mathcal{B}$. 
For this reason, in the analysis below we have $c^{z}_{jj}[0] > 0$ for every node $v_j$. 
Suppose that, following the random choices, nodes $v_1, v_2, v_3, v_4$ set
\begin{itemize}
\item[\ ] $v_1:$ \ \ $c^{y}_{11}[0] = 5$, $c^{y}_{21}[0] = 5$, $c^{y}_{31}[0] = 0$,
\item[\ ] \hspace{.73cm} $c^{z}_{11}[0] = 1$, $c^{z}_{21}[0] = 1$, $c^{z}_{31}[0] = 0$, 
\item[\ ]
\item[\ ] $v_2:$ \ \ $c^{y}_{22}[0] = 6$, $c^{y}_{42}[0] = 0$,
\item[\ ] \hspace{.73cm} $c^{z}_{22}[0] = 2$, $c^{z}_{42}[0] = 0$, 
\item[\ ]
\item[\ ] $v_3:$ \ \ $c^{y}_{13}[0] = 7$, $c^{y}_{23}[0] = 0$, $c^{y}_{33}[0] = 7$, 
\item[\ ] \hspace{.73cm} $c^{z}_{13}[0] = 1$, $c^{z}_{23}[0] = 0$, $c^{z}_{33}[0] = 1$, 
\item[\ ]
\item[\ ] $v_4:$ \ \ $c^{y}_{34}[0] = 2$, $c^{y}_{44}[0] = 2$, 
\item[\ ] \hspace{.73cm} $c^{z}_{34}[0] = 1$, $c^{z}_{44}[0] = 1$. 
\end{itemize}
We have that nodes $v_1$, $v_3$ and $v_4$ perform transmissions to nodes $v_2$, $v_1$ and $v_3$, respectively, whereas node $v_2$, transmits to itself. 

Then, each node $v_j$ receives from its in-neighbors $v_i \in \mathcal{N}_j^- \cup \{v_j\}$ the transmission variables $c^{y}_{ji}[0]$ and $c^{z}_{ji}[0]$ and, at time step $k=1$, it calculates its state variables $y^s_j[1]$, $z^s_j[1]$ and $q^s_j[1]$. 
The mass and state variables are shown in Table~\ref{tableProb_no_oscill} for $k=1$. 

\begin{center}
\captionof{table}{Mass and State Variables for Fig.~\ref{prob_example}}
\label{tableProb_no_oscill}
{\small 
\begin{tabular}{|c||c|c|c|c|c|}
\hline
Nodes &\multicolumn{5}{c|}{Mass and State Variables for $k=0$}\\
$v_j$ &$y_j[0]$&$z_j[0]$&$y^s_j[0]$&$z^s_j[0]$&$q^s_j[0]$\\
\cline{2-6}
 &  &  &  &  & \\
$v_1$ & 10 & 2 & 10 & 2 & 5\\
$v_2$ & 6 & 2 & 6 & 2 & 3\\
$v_3$ & 14 & 2 & 14 & 2 & 7\\
$v_4$ & 4 & 2 & 4 & 2 & 2\\
\hline
Nodes &\multicolumn{5}{c|}{Mass and State Variables for $k=1$}\\
$v_j$ &$y_j[1]$&$z_j[1]$&$y^s_j[1]$&$z^s_j[1]$&$q^s_j[1]$\\
\cline{2-6}
 &  &  &  &  & \\
$v_1$ & 12 & 2 & 12 & 2 & 6\\
$v_2$ & 11 & 3 & 11 & 3 & 3\\
$v_3$ & 9 & 2 & 9 & 2 & 4\\
$v_4$ & 2 & 1 & 2 & 1 & 2\\
\hline
\end{tabular}
}
\end{center}
\vspace{0.2cm}

Then, each node $v_j$ calculates and transmits its transmission variables $c^{y}_{lj}[k]$ and $c^{z}_{lj}[k]$ for every $v_l \in \mathcal{N}_j^+ \cup \{v_j\}$. 
Suppose that, following the random choices, nodes $v_1, v_2, v_3, v_4$ set
\begin{itemize}
\item[\ ] $v_1:$ \ \ $c^{y}_{11}[1] = 12$, $c^{y}_{21}[1] = 0$, $c^{y}_{31}[1] = 0$,
\item[\ ] \hspace{.73cm} $c^{z}_{11}[1] = 2$, $c^{z}_{21}[1] = 0$, $c^{z}_{31}[1] = 0$, 
\item[\ ]
\item[\ ] $v_2:$ \ \ $c^{y}_{22}[1] = 3$, $c^{y}_{42}[1] = 8$,
\item[\ ] \hspace{.73cm} $c^{z}_{22}[1] = 1$, $c^{z}_{42}[1] = 2$, 
\item[\ ]
\item[\ ] $v_3:$ \ \ $c^{y}_{13}[1] = 5$, $c^{y}_{23}[1] = 0$, $c^{y}_{33}[1] = 4$, 
\item[\ ] \hspace{.73cm} $c^{z}_{13}[1] = 1$, $c^{z}_{23}[1] = 0$, $c^{z}_{33}[1] = 1$, 
\item[\ ]
\item[\ ] $v_4:$ \ \ $c^{y}_{34}[1] = 2$, $c^{y}_{44}[1] = 0$, 
\item[\ ] \hspace{.73cm} $c^{z}_{34}[1] = 1$, $c^{z}_{44}[1] = 0$. 
\end{itemize}
In the analysis above we can see, for example, that node $v_3$ splits the value $y_3[1] = 9$ into $z_3[1] = 2$ almost equal pieces which are $4$ and $5$ respectively. 
Then, it keeps the piece $4$ (which has the minimum value) and transmits it to itself, and then it transmits the remaining piece $5$ to a randomly selected out-neighbor which happens to be node $v_1$.

In Fig.~\ref{example_plot_no_oscill} we plot the resulting state variable $q_j^s[k]$ of every node $v_j \in \mathcal{V}$, from which it can be seen that for $k \geq 9$ we have
$$
q^s_j[k] = \lfloor q \rfloor \ \ \text{or} \ \ q^s_j[k] = \lceil q \rceil , 
$$
for every $v_j \in \mathcal{V}$. 
This means that every node $v_j$ obtains, after a finite number of iterations, a quantized state $q_j^s$, which is equal either to the ceiling or to the floor of the real average $q$ of the initial states of the nodes. 
\end{example}

\begin{figure}[t]
\centering
\includegraphics[width=60mm]{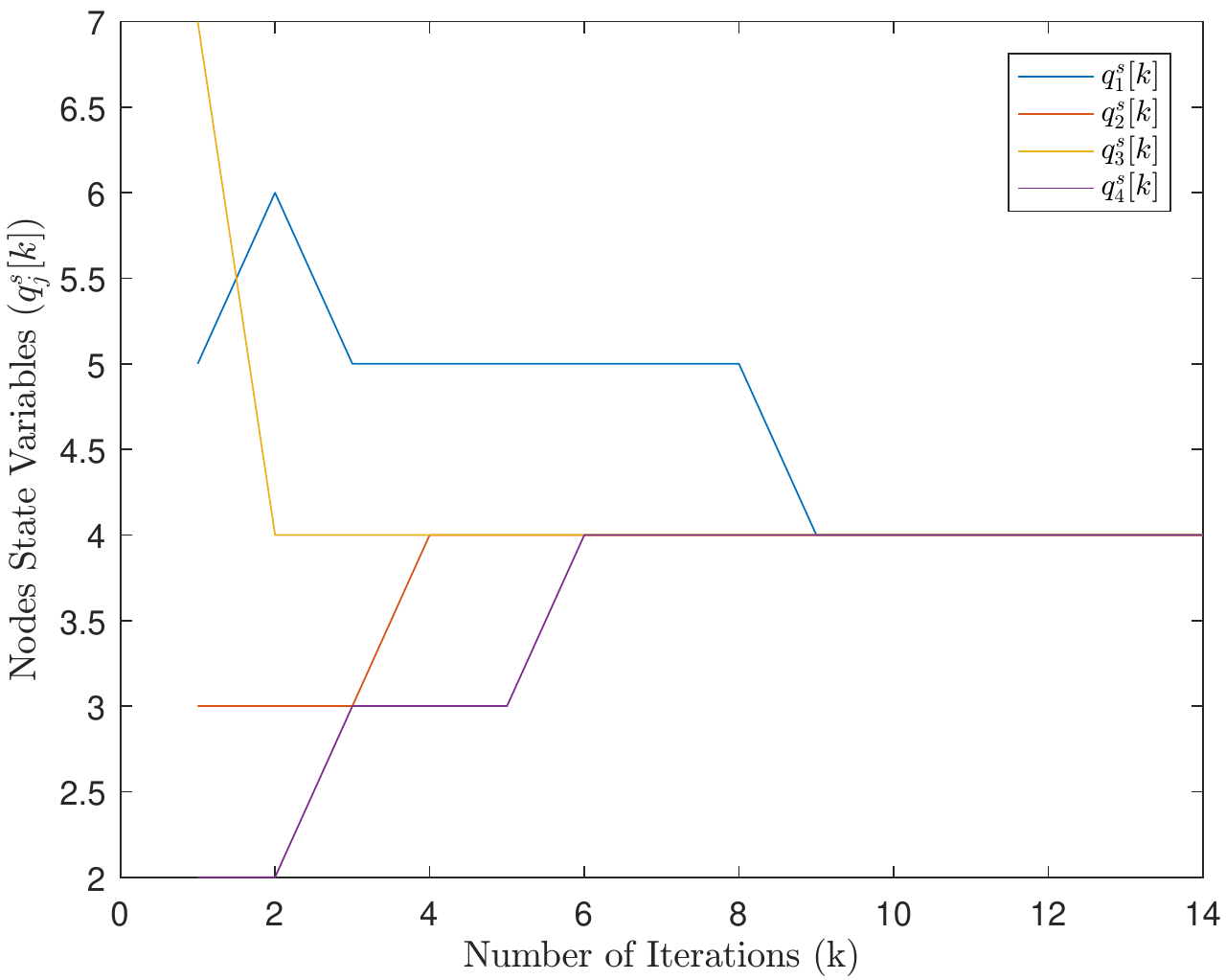}
\caption{Node state variables plotted against the number of iterations for Algorithm~\ref{algorithm_prob_no_oscil} for the digraph shown in Fig.~\ref{prob_example}.}
\label{example_plot_no_oscill}
\end{figure}

\begin{remark} 
It is important to note here that Algorithm~\ref{algorithm_prob_no_oscil} is significantly different from the algorithm in \cite{2020:RikosHadj_IFAC}. 
In \cite{2020:RikosHadj_IFAC} we have that (i) Initialization Step~$2$ is set to $z_j[0]=1$, (ii) Iteration Step~$1$ is set to $z_j[k] > 0$ and (iii) Iteration Step~$1.3$ is set to $mas^z[k] > 0$. 
This means that each node splits $y_j[k]$ into $z_j[k]$ equal pieces (or with maximum difference equal to $1$) and then transmits \textit{every} piece to randomly chosen out-neighbors or to itself. 
Steps (i), (ii), (iii) allow calculation of the quantized average in finite time, but they cause the state $q_j$ of each node $v_j$ to oscillate between $\lfloor q \rfloor$ and $\lceil q \rceil$ (where $q$ is defined in \eqref{real_av}) rather than stabilizing to one of the aforementioned values. 
During the operation of the proposed Algorithm~\ref{algorithm_prob_no_oscil}, steps (i), (ii) and (iii) are changed to $z_j[0]=2$, $z_j[k] > 1$, and $mas^z[k] > 1$, respectively. 
This means that each node splits $y_j[k]$ into $z_j[k]$ equal pieces (or with maximum difference equal to $1$), transmits one piece to itself and then transmits the remaining pieces to randomly chosen out-neighbors or to itself. 
These changes affect significantly the operation of the proposed algorithm and leads to fast convergence while avoiding oscillatory behavior of the agent's states. 
Specifically, by changing steps (i), (ii) and (iii), Algorithm~\ref{algorithm_prob_no_oscil} (a) achieves quantized average consensus in finite time, (b) maintains similar fast convergence speed as the algorithm in  \cite{2020:RikosHadj_IFAC} and (c) the state $q_j$ of each node $v_j$ stabilizes to $\lfloor q \rfloor$ or $\lceil q \rceil$. 
\\ 
Furthermore, Algorithm~\ref{algorithm_prob_no_oscil} is also significantly different from the algorithm in \cite{2020:Rikos_Quant_Cons}, where the authors presented a distributed deterministic algorithm in which every node $v_j$ ``merged'' (i.e., added) the incoming mass variables sent by its in-neighbours. 
No splitting was done and mass variables remained merged during the algorithm execution.
The authors of \cite{2020:Rikos_Quant_Cons} showed that every node $v_j$ calculated, after a finite number of time steps, a quantized fraction which is equal to the actual average $q$ of the initial states of the nodes (i.e., there was zero quantization error), but due to strict accumulation of the states, the proposed algorithm required a significant number of time steps and could also lead to a memory overflow problem if the initial node states were all close to the maximum representable value on the quantized scale (in which case the sum of those states may not be representable using a specific number of (fixed-point) bits).
By contrast, during the operation of Algorithm~\ref{algorithm_prob_no_oscil}, every node $v_j$ ``merges'' and then ``splits'' the incoming mass variables, sent by its in-neighbours; splitting of the mass variables allows faster convergence and avoids a memory overflow problem. 
As we will see in the following sections, this modification allows Algorithm~\ref{algorithm_prob_no_oscil} to significantly outperform (in terms of convergence speed) the algorithm presented in \cite{2020:Rikos_Quant_Cons} and other state-of-the-art algorithms in the available literature. 
\end{remark}

\subsection{Convergence of Algorithm~\ref{algorithm_prob_no_oscil}}\label{CONValgorithm_no_oscill}

We now show that, during the operation of Algorithm~\ref{algorithm_prob_no_oscil}, each agent $v_j$ addresses problem \textbf{(P1)} presented in Section~\ref{probForm} and reaches, after a finite number of time steps, a consensus state which is equal to either the ceiling or the floor of the actual average $q$ of the initial states of the nodes. 
We first consider Lemma~\ref{Lemma1_prob}, which is necessary for our subsequent development.

\begin{lemma}[\hspace{-0.00001cm}\cite{2020:RikosHadj_IFAC}]
\label{Lemma1_prob}
Consider a strongly connected digraph $\mathcal{G}_d = (\mathcal{V}, \mathcal{E})$ with $n=|\mathcal{V}|$ nodes and $m=|\mathcal{E}|$ edges.
Suppose that each node $v_j$ assigns a nonzero probability $b_{lj}$ to each of its outgoing edges $m_{lj}$, where $v_l \in \mathcal{N}^+_j \cup \{v_j\}$, as follows  
\begin{align*}
b_{lj} = \left\{ \begin{array}{ll}
         \frac{1}{1 + \mathcal{D}_j^+}, & \mbox{if $l = j$ or $v_{l} \in \mathcal{N}_j^+$,}\\
         0, & \mbox{if $l \neq j$ and $v_{l} \notin \mathcal{N}_j^+$.}\end{array} \right. 
\end{align*}
At time step $k=0$, node $v_j$ holds a ``token" while the other nodes $v_l \in \mathcal{V} - \{ v_j \}$ do not. 
Each node $v_j$ transmits the ``token'' (if it has it, otherwise it performs no transmission) according to the nonzero probability $b_{lj}$ it assigned to its outgoing edges $m_{lj}$. 
The probability $P^{n-1}_{T_i}$ that the token is at node $v_i$ after $n-1$ time steps satisfies 
$$
P^{n-1}_{T_i} \geq (1+\mathcal{D}^+_{max})^{-(n-1)} > 0 , 
$$
where $\mathcal{D}^+_{max} = \max_{v_j \in \mathcal{V}} \mathcal{D}^+_{j}$. 
\end{lemma}

\begin{theorem}\label{Theorem2_prob_no_oscill}
Consider a strongly connected digraph $\mathcal{G}_d = (\mathcal{V}, \mathcal{E})$ with $n=|\mathcal{V}|$ nodes and $m=|\mathcal{E}|$ edges and $z_j[0] = 1$ and $y_j[0] \in \mathbb{Z}$ for every node $v_j \in \mathcal{V}$ at time step $k=0$. 
Suppose that each node $v_j \in \mathcal{V}$ follows the Initialization and Iteration steps as described in Algorithm~\ref{algorithm_prob_no_oscil}. 
For any $\varepsilon$, where $0 < \varepsilon < 1$, there exists $k_0 \in \mathbb{Z}_+$, so that with probability $(1-\varepsilon)^{(y^{init}+n)}$ we have
$$ 
( q^s_j[k] = \lfloor q \rfloor \ \ \text{for} \ \ k \geq k_0) \ \ \text{or} \ \ ( q^s_j[k] = \lceil q \rceil \ \ \text{for} \ \ k \geq k_0)  ,
$$
for every $v_j \in \mathcal{V}$, where $q$ fulfills \eqref{real_av} and
\begin{align}
y^{init} & = & \sum_{\{v_j \in \mathcal{V}: y_j[0] > \lceil q \rceil\}} {(y_j[0] - \lceil q \rceil) } \ + \nonumber \\ 
 & & \sum_{\{v_j \in \mathcal{V}: y_j[0] < \lfloor q \rfloor\}} {(\lfloor q \rfloor - y_j[0])} , \label{initial_distance_no_oscill}
\end{align}
is the total initial state error (i.e., $y^{init}$ is the sum of the differences between (i) the value $\lceil q \rceil$ and the initial state $y_j[0]$ of each node $v_j$ that has an initial state higher than the ceiling of the real average and (ii) the value $\lfloor q \rfloor$ and the initial state $y_j[0]$ of each node $v_j$ that has an initial state less than the floor the real average). 
%\begin{equation}\label{initial_distance_no_oscill}
%y^{init} = \sum_{j=1}^n \max_{\{v_j \in \mathcal{V}: y_j[0] > \lceil q \rceil\}}{(y_j[0] - \lceil q \rceil)} + \sum_{j=1}^n \min_{\{v_j \in \mathcal{V}: y_j[0] < \lfloor q \rfloor\}}{(\lfloor q \rfloor - y_j[0])}.
%\end{equation}
\end{theorem}

\begin{pf}
The operation of Algorithm~\ref{algorithm_prob_no_oscil} can be interpreted as the ``random walk'' of $n$ ``tokens'' in a Markov chain, where $n=|\mathcal{V}|$. 
At each time step $k$, each of these $n$ tokens contains a pair of values $y[k]$, $z[k]$, for which $y[k] \in \mathbb{Z}$ and $z[k] = 1$. 
Specifically, at time step $k=0$, node $v_j$ holds two ``tokens". 
One token is $T_j^{ins}$ and is stationary, whereas the other token is $T_j^{out}$ and performs a random walk. 
Each token $T_j^{ins}$ and $T_j^{out}$ contains a pair of values $y_j^{ins}[k]$, $z_j^{ins}[k]$, and $y_j^{out}[k]$, $z_j^{out}[k]$, respectively. 
Initially, we have $y_j^{ins}[0] = y_j^{out}[0] = y_j[0]$ and $z_j^{ins}[0] = z_j^{out}[0] = z[0]$. 
At each time step $k$, each node $v_j$ keeps the token $T_j^{ins}$ (i.e., it never transmits it) while it transmits the token $T_j^{out}$, according to the nonzero probability $b_{lj}$ it assigned to its outgoing edges $m_{lj}$ during the Initialization Steps. 
If $v_j$ receives one or more tokens $T_i^{out}$ from its in-neighbors $v_i$ the values $y_i^{out}[k]$ and $y_j^{ins}[k]$ become equal (or with maximum difference equal to $1$); then $v_j$ transmits each received token $T_i^{out}$ to a randomly selected out-neighbor according to the nonzero probability $b_{lj}$ it assigned to its outgoing edges $m_{lj}$. 
Note here that during the operation of Algorithm~\ref{algorithm_prob_no_oscil} we have 
\begin{equation}\label{sum_preserve}
\sum_{j=1}^n y^{out}_j[k] + \sum_{j=1}^n y^{ins}_j[k] = 2 \sum_{j=1}^n y_j[0] , \ \forall k \in \mathbb{Z}_+ ,
\end{equation}
(i.e., the sum of the $y_j[k]$ values of the tokens at any given $k$ is equal to twice the initial sum). 

Let us now define 
\begin{equation}\label{BIG_lyap_no_oscil}
Y[k] = Y_1[k] + Y_2[k] ,
\end{equation} 
where 
\begin{equation}\label{lyap1_no_oscil}
Y_1[k] = \sum_{\{v_j \in \mathcal{V}: \lceil y_j[k]/z_j[k] \rceil > \lceil q \rceil\}} {(\lceil y_j[k]/z_j[k] \rceil - \lceil q \rceil)} ,
\end{equation} 
and
\begin{equation}\label{lyap2_no_oscil}
Y_2[k] = \sum_{\{v_j \in \mathcal{V}: \lfloor y_j[k]/z_j[k] \rfloor < \lfloor q \rfloor\}} {(\lfloor q \rfloor - \lfloor y_j[k]/z_j[k] \rfloor)} ,
\end{equation}
where $q$ satisfies \eqref{real_av}. 
We have that $Y_1[k]$ denotes the sum of the differences between the values $\lceil q \rceil$ and $y[k]$ of the tokens that have a $y$ value higher than the ceiling of the real average $\lceil q \rceil$, and $Y_2[k]$ denotes the sum of the differences between the values $\lfloor q \rfloor$ and $y[k]$ of the tokens that have $y$ value less than the floor the real average $\lfloor q \rfloor$.
Clearly, if $Y_1[k] > 0$ then at least one token has $y[k]$ value which is greater than the ceiling of the real average, and if $Y_2[k] > 0$ then at least one token has $y[k]$ value which is less than the floor of the real average, at time step $k$. 
From Iteration Steps~$1.3$ and $1.4$, we have that if two (or more) ``tokens'' $T_i^{out}$, $T_l^{out}$ (where $v_i, v_l \in \mathcal{V}$) meet at the same node $v_j$ with token $T_j^{ins}$ (which is kept in node $v_j$) during time step $k$, then their values $y[k]$ become equal (or with maximum difference equal to $1$). 
For the scenario $\lceil q \rceil > \lfloor q \rfloor$, we have at time step $k$ (note that similar arguments hold also for $\lceil q \rceil = \lfloor q \rfloor$): 
\\ \noindent
Case (i): If $Y_1[k] > 0$ and a token which has $y[k] > \lceil q \rceil$ meets with a token that has $y[k] \leq \lfloor q \rfloor$ then we have $Y_1[k+1] \leq Y_1[k] - 1$. 
\\ \noindent
Case (ii): If $Y_2[k] > 0$ and a token which has $y[k] < \lfloor q \rfloor$ meets with a token that has $y[k] \geq \lceil q \rceil$ then we have $Y_2[k+1] \leq Y_2[k] - 1$. 
\\ \noindent
Case (iii): If $Y_1[k] > 0$ and $Y_2[k] > 0$ and a token which has $y[k] > \lceil q \rceil$ meets with a token that has $y[k]  < \lfloor q \rfloor$ then we have $Y_1[k+1] \leq Y_1[k] - 1$ and $Y_2[k+1] \leq Y_2[k] - 1$. 
\\ \noindent 
Note that for the scenario $\lceil q \rceil = \lfloor q \rfloor$ we have that only Case~(iii) above holds. 
Case~(i) and Case~(ii) do not hold since the difference between the values $y[k]$ of the tokens that meet might be equal to one which means that the values of $Y_1[k]$ and $Y_2[k]$ will not decrease.

Clearly, we have 
$$
0 \leq Y[k+1] \leq Y[k] \leq y^{init} ,
$$ 
for all time steps $k$, where $y^{init}$ fulfills \eqref{initial_distance_no_oscill}. 
This means that if cases (i), (ii), (iii) hold for $y^{init}$ times the value of $Y$ becomes equal to zero (where $Y$ is defined in \eqref{BIG_lyap_no_oscil}). 
As a result, for every token the values $y$ become equal or have difference equal to one (recall that, during the operation of Algorithm~\ref{algorithm_prob_no_oscil}, we have that \eqref{sum_preserve} holds for every $k$).

In this proof, we focus on the scenario in which one token $y^{out}_i$ visits a specific node $v_j$ (for which it holds $| y^{out}_i - y^{ins}_j | > 1$) and obtains equal values (or with maximum difference between them equal to $1$) with the token $y^{ins}_j$ which is kept in node $v_j$. 
We consider and analyze the probability that this specific token visits this specific node in the network after a finite number of time steps. 
For any $\epsilon$, where $0 < \varepsilon < 1$, we show that (i) $\exists k_0' \in \mathbb{Z}_+$ for which with probability $(1-\varepsilon)^{y^{init}}$, it holds that $Y[k] = 0$ for every $k \geq k_0'$, and (ii) $\exists k_0 \in \mathbb{Z}_+$ for which, with probability $(1-\varepsilon)^{(y^{init}+n)}$, \eqref{alpha_q_no_oscill} holds for every $k \geq k_0$. 
This means that after a finite number of time steps $k_0$ the value $y[k]$ of every token is equal either to $\lfloor q \rfloor$ or to $\lceil q \rceil$, and for the state variable $q^s_j[k]$ of every node $v_j$ we have $ q^s_j[k] = \lfloor q \rfloor $ or $ q^s_j[k] = \lceil q \rceil $, respectively.

Let us consider tokens $T_{\lambda}^{out}$ and $T_{i}^{ins}$ for which it holds (i) $y_{\lambda}^{out} \geq \lceil q \rceil$, $y_{i}^{ins} < \lfloor q \rfloor$, or (ii) $y_{\lambda}^{out} > \lceil q \rceil$, $y_{i}^{ins} \leq \lfloor q \rfloor$, or (iii) $y_{\lambda}^{out} < \lfloor q \rfloor$, $y_{i}^{ins} \geq \lceil q \rceil$, or (iv) $y_{\lambda}^{out} \leq \lfloor q \rfloor$, $y_{i}^{ins} > \lceil q \rceil$.  
During the operation of Algorithm~\ref{algorithm_prob_no_oscil}, $n$ ``tokens'' perform {\em independent} random walks, and from Lemma~\ref{Lemma1_prob} we have that the probability $P^{n-1}_{T^{out}}$ that ``the specific token $T_{\lambda}^{out}$ is at node $v_i$ after $n-1$ time steps'' is 
\begin{equation}\label{lowerProf_no_oscil}
P^{n-1}_{T^{out}} \geq (1+\mathcal{D}^+_{max})^{-(n-1)} .
\end{equation}
This means that the probability $P^{n-1}_{N\_ T^{out}}$ that ``the specific token $T_{\lambda}^{out}$ has not visited node $v_i$ after $n-1$ time steps'' is
\begin{equation}\label{lowerProf_not_no_oscil}
P^{n-1}_{N\_ T^{out}} \leq 1- (1+\mathcal{D}^+_{max})^{-(n-1)} .
\end{equation}
By extending this analysis, we can state that for any $\epsilon$, where $0 < \varepsilon < 1$ and after $\tau(n-1)$ time steps where
\begin{equation}\label{windows_for_conv_1_no_oscil}
\tau \geq \Big \lceil \dfrac{\log{\epsilon}}{\log{(1 - (1+\mathcal{D}^+_{max})^{-(n-1)})}} \Big \rceil ,
\end{equation}
the probability $P^{\tau}_{N\_ T^{out}}$ that ``the specific token $T_{\lambda}^{out}$ has not visited node $v_i$ after $\tau (n-1)$ time steps'' is
\begin{equation}\label{ProbNotMeet_after_t_no_oscil}
P^{\tau}_{N\_ T^{out}} \leq [P^{n-1}_{N\_ T^{out}}]^{\tau} \leq \epsilon .
\end{equation}
This means that after $\tau (n-1)$ time steps, where $\tau$ fulfills \eqref{windows_for_conv_1_no_oscil}, the probability that ``the specific token $T_{\lambda}^{out}$ has visited node $v_i$ after $\tau (n-1)$ time steps'' is equal to $1-\epsilon$.

As a result, after $\tau (n-1)$ time steps, where $\tau$ fulfills \eqref{windows_for_conv_1_no_oscil}, we have that if $Y_1[k] > 0$ and/or $Y_2[k] > 0$ at time step $k$, then it holds that $Y_1[k + \tau (n-1)] \leq Y_1[k] - 1$ and/or $Y_2[k + \tau (n-1)] \leq Y_2[k] - 1$ with probability $1-\varepsilon$. 
By extending this analysis, we have that for $k \geq y^{init} \tau (n-1)$ time steps, where $y^{init}$ is given by \eqref{initial_distance_no_oscill}, we have $Y[k] = 0$ with probability $( 1-\varepsilon )^{y^{init}}$. 
Therefore for $k \geq y^{init} \tau (n-1)$, we have that the value $y[k]$ of every token is equal to either $\lfloor q \rfloor$ or $\lceil q \rceil$ with probability $( 1-\varepsilon )^{y^{init}}$. 
Since \eqref{sum_preserve} holds, for $k \geq y^{init} \tau (n-1)$ we have 
\begin{equation}\label{y_out_conv_no_oscil}
\lfloor y_j^{out}[k] / z_j^{out}[k] \rfloor = \lfloor q \rfloor \ \ \text{or} \ \ \lceil y_j^{out}[k] / z_j^{out}[k] \rceil = \lceil q \rceil ,
\end{equation}
and 
\begin{equation}\label{y_ins_conv_no_oscil}
\lfloor y_j^{ins}[k] / z_j^{ins}[k] \rfloor = \lfloor q \rfloor \ \ \text{or} \ \ \lceil y_j^{ins}[k] / z_j^{ins}[k] \rceil = \lceil q \rceil ,
\end{equation}
for every $v_j \in \mathcal{V}$ with probability $( 1-\varepsilon )^{y^{init}}$. 
Furthermore, we have that for $k \geq y^{init} \tau (n-1)$ it holds 
\begin{eqnarray}\label{number_of_aver_tokens}
| \{ T_j^{ins}, v_j \in \mathcal{V} | y_j^{ins}[k] = \lfloor q \rfloor \} | & + & \nonumber \\
| \{ T_j^{out}, v_j \in \mathcal{V} | y_j^{out}[k] = \lfloor q \rfloor \} | & = & 2n - 2R .  \;
\end{eqnarray}
with probability $( 1-\varepsilon )^{y^{init}}$, where $| \{ T_j^{ins}, v_j \in \mathcal{V} | y_j^{ins}[k] = \lfloor q \rfloor \} |$ is the cardinality of the set of tokens $T_j^{ins}$ which have $y_j^{ins}$ value equal to $\lfloor q \rfloor$, $| \{ T_j^{out}, v_j \in \mathcal{V} | y_j^{out}[k] = \lfloor q \rfloor \} |$ is the cardinality of the set of tokens $T_j^{out}$ which have $y_j^{out}$ value equal to $\lfloor q \rfloor$ and $R$ is defined in \eqref{equation_mod}. 
This means that the number of tokens with $y_j$ value equal to $\lfloor q \rfloor$ is $2n - 2R$.

Continuing the analysis, let us consider now the following two cases
\begin{enumerate}
\item $2n - 2R \geq n$ (or $R < n/2$),
\item $2n - 2R < n$ (or $R > n/2$),
\end{enumerate}
where $n$ is the number of nodes and $R$ is defined in \eqref{equation_mod}. 

For the first case, we have that the number of tokens which have value equal to $\lfloor q \rfloor$ is greater than (or equal to) the number of nodes. 
This means that by executing Algorithm~\ref{algorithm_prob_no_oscil} for an additional number of $n \tau (n-1)$ time steps, where $\tau$ fulfills \eqref{windows_for_conv_1_no_oscil}, we have that every node will receive at least one token with value $\lfloor q \rfloor$ with probability $(1 - \varepsilon)^n$. 
[The reason is that during the first $\tau(n-1)$ steps, one of the tokens with value $\lfloor q \rfloor$ will reach node $v_1$ with probability $1 - \varepsilon$; during the second $\tau(n-1)$ steps, one of the tokens with value $\lfloor q \rfloor$ will reach node $v_2$ with probability $1 - \varepsilon$, and so on. 
During the last $\tau(n-1)$ steps, one of the tokens with value $\lfloor q \rfloor$ will reach node $v_n$ with probability $1 - \varepsilon$.] 
From Iteration Steps~$1.3$ and $1.4$ of Algorithm~\ref{algorithm_prob_no_oscil} we have that if node $v_j$ receives a token with value $\lfloor q \rfloor$, then the value of its $y_j^{ins}$ token becomes equal to $\lfloor q \rfloor$ which means that also the value of its state variable $q^s_j$ becomes equal to $\lfloor q \rfloor$. 
As a result, since $2n - 2R \geq n$, for $k \geq (y^{init} + n) \tau (n-1)$, where $y^{init}$ fulfills \eqref{initial_distance_no_oscill} and $\tau$ fulfills \eqref{windows_for_conv_1_no_oscil}, we have 
$$
y^{ins}_j[k] = \lfloor q \rfloor, \ \text{for every} \ v_j \in \mathcal{V} , 
$$
and 
$$
q^{s}_j[k] = \lfloor q \rfloor, \ \text{for every} \ v_j \in \mathcal{V} , 
$$
with probability $(1-\varepsilon)^{(y^{init} + n)}$.

For the second case, the number of tokens which have value equal to $\lfloor q \rfloor$ is less than the number of nodes. 
Identically to the first case, by executing Algorithm~\ref{algorithm_prob_no_oscil} for an additional number of $n \tau (n-1)$ time steps, where $\tau$ fulfills \eqref{windows_for_conv_1_no_oscil}, for $k \geq (y^{init} + n) \tau (n-1)$ we have with probability $(1-\varepsilon)^{(y^{init} + n)}$ that 
$$
y^{ins}_j[k] = \lfloor q \rfloor \ \text{and} \ q^{s}_j[k] = \lfloor q \rfloor, \ \text{for every} \ v_j \in \mathcal{V'} , 
$$
where $\mathcal{V'} \subset \mathcal{V}$ and $| \mathcal{V'} | = 2n - 2R$, and 
$$
y^{ins}_j[k] = \lceil q \rceil \ \text{and} \ q^{s}_j[k] = \lceil q \rceil, \ \text{for every} \ v_j \in \mathcal{V''} , 
$$
where $\mathcal{V''} \subset \mathcal{V}$ and $| \mathcal{V''} | = 2R - n$.

As a result, during the operation of Algorithm~\ref{algorithm_prob_no_oscil}, for $k \geq (y^{init}+n) \tau (n-1)$ we have 
$$
q^s_j[k] = \lfloor q \rfloor \ \ \text{or} \ \ q^s_j[k] = \lceil q \rceil  ,
$$
for every $v_j \in \mathcal{V}$, with probability $(1-\varepsilon)^{(y^{init} + n)}$.   $ \hfill  \qed $
\end{pf}

\begin{remark}
Note that the we can also bound the number of time steps $k_0$ needed for Algorithm~\ref{algorithm_prob_no_oscil} to converge according to a desired probability $p_0$. 
From the proof of Theorem~\ref{Theorem2_prob_no_oscill}, we have that each node $v_j$ can calculate the quantized average of the initial states with probability $p_0 \geq (1-\varepsilon)^{(y^{init}+n)}$, where $0 < \varepsilon < 1$ and $y^{init}$ fulfills \eqref{initial_distance_no_oscill}, after at least $k_0 \geq (y^{init} + n) \tau (n-1)$ time steps, where $y^{init}$ fulfills \eqref{initial_distance_no_oscill} and $\tau$ fulfills \eqref{windows_for_conv_1_no_oscil}. 
This means that if we want to guarantee convergence of Algorithm~\ref{algorithm_prob_no_oscil} with probability greater than (or equal to) $p_0$, we have to choose $\varepsilon$ for which it holds $\varepsilon \geq 1 - 2^{\frac{\log_2{p_0}}{y^{init} + n}}$. 
Then, according to the proof of Theorem~\ref{Theorem2_prob_no_oscill}, we need to execute Algorithm~\ref{algorithm_prob_no_oscil} for a number of time steps $k_0$ for which it holds 
$$
k_0 \geq (y^{init}+n) \frac{\log_2{\Bigl ( 1 - 2^{\frac{\log_2{p_0}}{y^{init} + n}} \Bigr )}}{\log_2{\Bigl ( 1 - (1+\mathcal{D}^+_{max})^{-(n-1)} \Bigr )}} (n-1), 
$$
where $\mathcal{D}^+_{max} = \max_{v_j \in \mathcal{V}} \mathcal{D}^+_{j}$ and $y^{init}$ fulfills \eqref{initial_distance_no_oscill}. 
\end{remark}

\begin{remark}
Algorithm~\ref{algorithm_prob_no_oscil} possesses attractive features for consensus-based distributed optimization. 
Apart from operating with quantized states which reduces the communication bottleneck \cite{2020:Jadbabaie, 2020:Shlezinger}, it also allows for fast distributed averaging, which makes it suitable as an intermediate step between the optimization operations \cite{2020:Khatana_Salapaka, 2020:Themis_Kalyvianaki}. 
In the latter case, the convergence speed of the averaging algorithm plays a significant role for the overall convergence speed of the optimization procedure (as we will see in subsequent sections, the convergence speed of Algorithm~\ref{algorithm_prob_no_oscil} significantly outperforms state-of-the-art algorithms in the available literature). 
Algorithm~\ref{algorithm_prob_no_oscil} can also find various applications on load balancing and voting schemes where each node needs to calculate a specific state rather than oscillate between two different states/decisions. 
\end{remark}

\section{Quantized Averaging over Dynamic Digraphs}\label{algorithm_no_oscil_dyn_digr}

In this section,we present a distributed algorithm which addresses problem \textbf{(P2)} presented in Section~\ref{probForm}. 
This means that the results of Section~\ref{algorithm_no_oscil} are extended to include directed topologies with time-varying communication links.  
We assume that, at each time step $k$, the interconnections between components in the multi-component system are captured by a digraph $\mathcal{G}_d[k] = (\mathcal{V}, \mathcal{E}[k])$ in which the set of nodes is fixed but the communication links may change.

\begin{assum}
\ 
\begin{enumerate}
\item[$A_1.$] At each time step $k$, each node $v_j$ has knowledge of the set of its out-neighbors $\mathcal{N}_j^+[k]$ and the number of its out-neighbors $\mathcal{D}_j^+[k]$. 
\item[$A_2.$] For any infinite sequence of dynamic digraphs $\mathcal{G}_d[1]$, $\mathcal{G}_d[2]$, ..., $\mathcal{G}_d[k]$, ..., there is a finite window length $l \in \mathbb{N}$ and an infinite sequence of time instants $t_0$, $t_1$, ..., $t_m$, ..., where $t_0 = 0$, such that for any $m \in \mathbb{Z}_+$, we have $0 < t_{m+1} - t_m < l < \infty$ and the union graph $\mathcal{G}^{t_{m}, ..., t_{m+1}-1}_d$, is equal to the nominal digraph $\mathcal{G}_d$ which is assumed to be strongly connected. 
\item[$\overline{A_2}$] There is a finite collection of dynamic digraphs $\mathcal{G}_{d_1}$, $\mathcal{G}_{d_2}$, ..., $\mathcal{G}_{d_M}$, such that the union graph is strongly connected and at each time step $k$ one such topology is selected independently in an i.i.d. manner. 
Specifically, at time step $k$, we have $G_d[k] = G_{d_{\theta}}$ for some $\theta \in \{1, 2, ..., M\}$ with probability $p_\theta$ where $\sum_{\theta = 1}^M p_\theta = 1$.
\end{enumerate}
\end{assum}

Assumption $A_1$ implies that the transmitting node knows the number of nodes it transmits messages to at each time instant. 
In an undirected graph setting, this is not difficult and can be done straightforwardly; in a directed graph setting, this is challenging but there are ways in which knowledge of the out-degree might be possible. 
For example, there can be an acknowledgement signal via a \textit{distress signal} (special tone in a control slot or some separate control channel) sent at higher power than normal so that it is received by transmitters in its vicinity \cite{2020:Bambos}. 
Knowledge of the out-degree is also possible if the nodes periodically perform checks to determine the number of their out-neighbors (e.g., by periodically transmitting the distress signals mentioned above). 
\\ \noindent
Assumption $A_2$ (or $\overline{A_2})$ is sufficient for the existence of at least one directed path between any pair of nodes infinitely often.

Under the above assumptions, during the operation of Algorithm~\ref{algorithm_prob_no_oscil_dynamic}, each node $v_j$ is required to calculate the nonzero probabilities $b_{lj}[k]$ for each of its outgoing edges $m_{lj}[k]$ (where $v_l \in \mathcal{N}^+_j[k] \cup \{v_j\}$) at each time step $k$. 
This calculation is due to the dynamic nature of the communication topology $\mathcal{G}_d[k]$. 
Note that since each transmitting node $v_j$ has instant knowledge of its out-degree then it sets the weights $b_{lj}[k]$ to be equal to $b_{lj}[k] = \frac{1}{1 + \mathcal{D}_j^+[k]}$ for $v_{l} \in \mathcal{N}_j^+[k] \cup \{ v_j \}$. 
This choice satisfies $\sum_{l=1}^{n} b_{lj}[k] = 1$ for all $v_j \in \mathcal{V}$ which means that the transition matrix $\mathcal{B}[k] = [b_{lj}[k]]$ is column-stochastic at every time step $k$. 
Furthermore, unspecified weights in $\mathcal{B}[k]$ are set to zero and correspond to pairs of nodes $(v_l, v_j)$ that are not connected at time step $k$, i.e., $b_{lj}[k] = 0$ for $v_{l} \notin \mathcal{N}_j^+[k] \cup \{ v_j \}$.

%the operation of each node $v_j$ (that is described by Algorithm~\ref{algorithm_prob_no_oscil_dynamic}) which is similar to Algorithm~\ref{algorithm_prob_no_oscil}, with the main difference being that, due to the dynamic nature of the communication topology $\mathcal{G}_d[k]$, each node $v_j$ calculates the nonzero probabilities $b_{lj}[k]$ for each of its outgoing edges $m_{lj}[k]$ (where $v_l \in \mathcal{N}^+_j[k] \cup \{v_j\}$) at each time step $k$. 
%The subsequent operation of Algorithm~\ref{algorithm_prob_no_oscil_dynamic} is similar to Algorithm~\ref{algorithm_prob_no_oscil} but each step is adjusted according to the time-varying sets of in- and out-neighbors, $\mathcal{N}^-_j[k]$ and $\mathcal{N}^+_j[k]$. 

\begin{varalgorithm}{2}
\caption{Quantized Average Consensus Over Dynamic Digraphs}
\noindent \textbf{Input} A set of digraphs $\mathcal{G}_d[k] = (\mathcal{V}, \mathcal{E}[k])$ with $n=|\mathcal{V}|$ nodes and $m[k]=|\mathcal{E}[k]|$ edges, for each $k=0,1,2,\dots$ . 
Each node $v_j\in \mathcal{V}$ has an initial state $y_j[0] \in \mathbb{Z}$. \\
\textbf{Initialization:} Each node $v_j \in \mathcal{V}_p$ sets $y_j[0] := 2y_j[0]$, $z_j[0] = 2$. \\
\textbf{Iteration:} For $k=0,1,2,\dots$, each node $v_j \in \mathcal{V}_p$ does the following:
\begin{list4}
\item[$1)$] assigns a nonzero probability $b_{lj}[k]$ to each of its outgoing edges $m_{lj}[k]$, where $v_l \in \mathcal{N}^+_j[k] \cup \{v_j\}$, as follows  
\begin{align*}
b_{lj}[k] = \left\{ \begin{array}{ll}
         \frac{1}{1 + \mathcal{D}_j^+[k]}, & \mbox{if $l = j$ or $v_{l} \in \mathcal{N}_j^+[k]$,}\\
         0, & \mbox{if $l \neq j$ and $v_{l} \notin \mathcal{N}_j^+[k]$.}\end{array} \right.
\end{align*} 
\item[$2)$] \textbf{if} $z_j[k] > 1$, \textbf{then} 
\begin{list4a}
\item[$2.1)$] sets $z^s_j[k] = z_j[k]$, $y^s_j[k] = y_j[k]$, 
$
q^s_j[k] = \Bigl \lfloor \frac{y^s_j[k]}{z^s_j[k]} \Bigr \rfloor \ ;
$
\item[$2.2)$] sets (i) $mas^y[k] = y_j[k]$, $mas^z[k] = z_j[k]$; (ii) $c^{y}_{lj}[k] = 0$, $c^{z}_{lj}[k] = 0$, for every $v_l \in \mathcal{N}^+_j[k] \cup \{v_j\}$; (iii) $\delta = \lfloor mas^y[k] / mas^z[k] \rfloor$, $mas^{rem}[k]= y_j[k] - \delta \ mas^z[k]$; 
\item[$2.3)$] \textbf{while} $mas^z[k] > 1$, \textbf{then}
\begin{list4a}
\item[$2.3a)$] chooses $v_l \in \mathcal{N}^+_j[k] \cup \{v_j\}$ randomly according to $b_{lj}$; 
\item[$2.3b)$] sets (i) $c^{z}_{lj}[k] := c^{z}_{lj}[k] + 1$, $c^{y}_{lj}[k] := c^{y}_{lj}[k] + \delta$; (ii) $mas^z[k] := mas^z[k] - 1$, $mas^y[k] := mas^y[k] - \delta$;
\item[$2.3c)$] \textbf{if} $mas^{rem}[k] > 0$, \textbf{then} sets $c^{y}_{lj}[k] := c^{y}_{lj}[k] + 1$, $mas^{rem}[k] := mas^{rem}[k]- 1$. 
\end{list4a}
\item[$2.4)$] sets $c^{y}_{jj}[k] := c^{y}_{jj}[k] + mas^y[k]$, $c^{z}_{jj}[k] := c^{z}_{jj}[k] + mas^z[k]$; 
\item[$2.5)$] \textbf{if} $c^{z}_{lj}[k] > 0$, \textbf{then} transmits $c^{y}_{lj}[k]$, $c^{z}_{lj}[k]$ to out-neighbor $v_l$, for every $v_l \in \mathcal{N}^+_j[k]$.
\end{list4a}
\item \textbf{else if} $z_j[k] \leq 1$, \textbf{then} sets $c^{y}_{jj}[k] = y[k]$, $c^{z}_{jj}[k] = z[k]$; 
\item[$3)$] receives $c^{y}_{ji}[k]$, $c^{z}_{ji}[k]$ from $v_i \in \mathcal{N}_j^-[k]$ and updates $y_j[k+1]$, $z_j[k+1]$ according to \eqref{no_del_eq_y_no_oscil}, \eqref{no_del_eq_z_no_oscil}. 
\end{list4}
\textbf{Output:} \eqref{alpha_q_no_oscill} holds for every $v_j \in \mathcal{V}$. 
\label{algorithm_prob_no_oscil_dynamic}
\end{varalgorithm}

\begin{remark}
%Note here that Algorithm~\ref{algorithm_prob_no_oscil_dynamic} along with the convergence proof presented below becomes identical to Algorithm~\ref{algorithm_prob_no_oscil} and its convergence proof when we have static directed communication topologies. 
Note here that the operation of Algorithm~\ref{algorithm_prob_no_oscil} is simpler than Algorithm~\ref{algorithm_prob_no_oscil_dynamic} since the computation of the nonzero probabilities $b_{lj}$ (for each of its outgoing edges $m_{lj}$) is done by each node $v_j$ only during the initialization steps and is not repeated at each time step $k$. 
This means that for static directed communication topologies, both algorithms will exhibit the same performance and convergence rate but, due to their operation, Algorithm~\ref{algorithm_prob_no_oscil} will consume less computational resources compared to Algorithm~\ref{algorithm_prob_no_oscil_dynamic}. 
\end{remark}

\subsection{Convergence of Algorithm~\ref{algorithm_prob_no_oscil_dynamic}}\label{CONValgorithm_no_oscill_dynamic}

We now show that, during the operation of Algorithm~\ref{algorithm_prob_no_oscil_dynamic}, each agent $v_j$ reaches a consensus state which is equal either to the ceiling or the floor of the actual average $q$ of the initial states of the nodes for the case where the digraph $\mathcal{G}_d[k]$ is dynamic (i.e., each agent addresses problem \textbf{(P2)} presented in Section~\ref{probForm}). 
We analyze the operation of Algorithm~\ref{algorithm_prob_no_oscil_dynamic} considering the set of assumptions $A_1, A_2$ (the set of assumptions $A_1, \overline{A_2}$ can be proven identically).

\begin{theorem}\label{Theorem2_prob_no_oscill_dynamic}
Consider a sequence of digraphs $\mathcal{G}_d[k] = (\mathcal{V}, \mathcal{E}[k])$, $k=0, 1, 2, ...$, with $n=|\mathcal{V}|$ nodes, $m[k]=|\mathcal{E}[k]|$ edges so that assumptions $A_1, A_2$ hold for $\mathcal{G}_d[k]$ over all $k$.  
Every node $v_j \in \mathcal{V}$ has the variables $z_j[0] = 1$ and $y_j[0] \in \mathbb{Z}$ at time step $k=0$, and it follows the Initialization and Iteration steps as described in Algorithm~\ref{algorithm_prob_no_oscil_dynamic}. 
For any $\varepsilon$, where $0 < \varepsilon < 1$, there exists $k_0 \in \mathbb{Z}_+$, so that with probability $(1-\varepsilon)^{(y^{init}+n)}$ we have
$$ 
( q^s_j[k] = \lfloor q \rfloor \ \ \text{for} \ \ k \geq k_0 ) \ \ \text{or} \ \ ( q^s_j[k] = \lceil q \rceil \ \ \text{for} \ \ k \geq k_0)  ,
$$
for every $v_j \in \mathcal{V}$, where $y^{init}$ is given in \eqref{initial_distance_no_oscill} and $q$ is given by \eqref{real_av} (i.e., for $k \geq k_0$ every node $v_j$ has calculated the ceiling or the floor of the actual average $q$ of the initial states). 
\end{theorem}

\begin{pf}
The operation of Algorithm~\ref{algorithm_prob_no_oscil_dynamic} can be interpreted as the ``random walk'' of $n$ ``tokens'' in a \textit{dynamic} (inhomogeneous) Markov chain (i.e., interconnections change over time) with $n=|\mathcal{V}|$ states. 
Each node $v_j$ at time step $k=0$ holds two ``tokens" $T_j^{ins}$ (which is stationary) and $T_j^{out}$ (which performs a random walk) and they each contain a pair of values $y_j^{ins}[k]$, $z_j^{ins}[k]$, and $y_j^{out}[k]$, $z_j^{out}[k]$, respectively, for which it holds that $y_j^{ins}[0] = y_j^{out}[0] = y_j[0] \in \mathbb{Z}$ and $z_j^{ins}[0] = z_j^{out}[0] = z[0] = 1$. 
At each time step $k$, each node $v_j$ keeps the token $T_j^{ins}$ (i.e., it never transmits it) while it transmits the token $T_j^{out}$, according to the nonzero probability $b_{lj}[k]$ it assigned to its outgoing edges $m_{lj}[k]$ during time step $k$. 
If $v_j$ receives one or more tokens $T_i^{out}$ from its in-neighbors the values $y_i^{out}[k]$ and $y_j^{ins}[k]$ become equal (or with maximum difference equal to $1$); then $v_j$ transmits each received token $T_i^{out}$ to a randomly selected out-neighbor according to the nonzero probability $b_{lj}[k]$. 
Note here that \eqref{sum_preserve} holds during the operation of Algorithm~\ref{algorithm_prob_no_oscil_dynamic} which means the sum of the $y_j[k]$ values of the tokens at any given $k$ is equal to twice the initial sum. 

Let us now define equations $Y[k]$, $Y_1[k]$, $Y_2[k]$ from \eqref{BIG_lyap_no_oscil}, \eqref{lyap1_no_oscil}, \eqref{lyap2_no_oscil}, respectively, where $Y_1[k]$, $Y_2[k]$ denote the sum of the differences between the values $y[k]$ and $\lceil q \rceil$ of the tokens that have a $y$ value higher than the ceiling of the real average $\lceil q \rceil$, and the sum of the differences between the values $y[k]$ and $\lfloor q \rfloor$ of the tokens that have $y$ value less than the floor of the real average $\lfloor q \rfloor$, respectively. 
From Iteration Steps~$2.3$ and $2.4$, we have that if two (or more) ``tokens'' $T_i^{out}$, $T_l^{out}$ (where $v_i, v_l \in \mathcal{V}$) meet at the same node $v_j$ with token $T_j^{ins}$ during time step $k$, then their values $y[k]$ become equal (or with maximum difference equal to $1$).
For the scenario $\lceil q \rceil > \lfloor q \rfloor$, we have at time step $k$ (note that similar arguments hold also for $\lceil q \rceil = \lfloor q \rfloor$): 
\\ \noindent
Case (i): If $Y_1[k] > 0$ and a token which has $y[k] > \lceil q \rceil$ meets with a token that has $y[k] \leq \lfloor q \rfloor$ then we have $Y_1[k+1] \leq Y_1[k] - 1$. 
\\ \noindent
Case (ii): If $Y_2[k] > 0$ and a token which has $y[k] < \lfloor q \rfloor$ meets with a token that has $y[k] \geq \lceil q \rceil$ then we have $Y_2[k+1] \leq Y_2[k] - 1$. 
\\ \noindent
Case (iii): If $Y_1[k] > 0$ and $Y_2[k] > 0$ and a token which has $y[k] > \lceil q \rceil$ meets with a token that has $y[k]  < \lfloor q \rfloor$ then we have $Y_1[k+1] \leq Y_1[k] - 1$ and $Y_2[k+1] \leq Y_2[k] - 1$. 
\\ \noindent 
Note that for the scenario $\lceil q \rceil = \lfloor q \rfloor$ we have that only Case~(iii) above holds. 
Case~(i) and Case~(ii) do not hold since the difference between the values $y[k]$ of the tokens that meet might be equal to one which means that the values of $Y_1[k]$ and $Y_2[k]$ will not decrease.

Clearly we have 
$$
0 \leq Y[k+1] \leq Y[k] \leq y^{init} ,
$$ 
for all time steps $k$, where $y^{init}$ fulfills \eqref{initial_distance_no_oscill}. 
This means that if cases (i), (ii), (iii) hold $y^{init}$ times the value of $Y$ becomes equal to zero (where $Y$ is defined in \eqref{BIG_lyap_no_oscil}). 
As a result, for every token the values $y$ become equal or have difference equal to one (recall that, during the operation of Algorithm~\ref{algorithm_prob_no_oscil_dynamic}, we also have that \eqref{sum_preserve} holds for every $k$).

In this proof, we consider and analyze the probability that a specific token, with value $y^{out}_i$, visits a specific node $v_j$, with token value $y^{ins}_j$, in the network after a finite number of time steps and obtains equal values (or with maximum difference between them equal to $1$) with the token $y^{ins}_j$, where for tokens  $T_{i}^{out}$ and $T_{j}^{ins}$ it holds (i) $y_{\lambda}^{out} \geq \lceil q \rceil$, $y_{i}^{ins} < \lfloor q \rfloor$, or (ii) $y_{\lambda}^{out} > \lceil q \rceil$, $y_{i}^{ins} \leq \lfloor q \rfloor$, or (iii) $y_{\lambda}^{out} < \lfloor q \rfloor$, $y_{i}^{ins} \geq \lceil q \rceil$, or (iv) $y_{\lambda}^{out} \leq \lfloor q \rfloor$, $y_{i}^{ins} > \lceil q \rceil$. 
Note that the main difference with the proof of Theorem~\ref{Theorem2_prob_no_oscill} is that during the operation of Algorithm~\ref{algorithm_prob_no_oscil_dynamic} the underlying communication topology is dynamic (i.e., interconnections change at every time step $k$). 
For any $\epsilon$, where $0 < \varepsilon < 1$, we show that (i) $\exists k_0' \in \mathbb{Z}_+$ for which with probability $(1-\varepsilon)^{y^{init}}$, it holds that $Y_1[k] = 0$ and $Y_2[k] = 0$ for every $k \geq k_0'$, and (ii) $\exists k_0 \in \mathbb{Z}_+$ for which  \eqref{alpha_q_no_oscill} holds with probability $(1-\varepsilon)^{(y^{init}+1)}$, for every $k \geq k_0$. 
This means that after a finite number of time steps $k_0$ the value $y[k]$ of every token is equal either to $\lfloor q \rfloor$ or to $\lceil q \rceil$, and for the state variable $q^s_j[k]$ of every node $v_j$ we have $ q^s_j[k] = \lfloor q \rfloor $ or $ q^s_j[k] = \lceil q \rceil $, respectively.

Let us consider tokens $T_{\lambda}^{out}$ and $T_{i}^{ins}$ for which it holds $| y_{\lambda}^{out} - y_{i}^{ins} | > 1$. 
During the operation of Algorithm~\ref{algorithm_prob_no_oscil_dynamic}, $n$ ``tokens'' perform {\em independent} random walks over a dynamic digraph $\mathcal{G}_d[k]$.

Since $b_{lj}[k] \geq (1+\mathcal{D}^+_{max})^{-1}$ (where $\mathcal{D}^+_{max}$ is defined in Lemma~\ref{Lemma1_prob} for the nominal digraph $\mathcal{G}_d$) and assumptions $A_1, A_2$ hold for $\mathcal{G}_d[k]$ during all $k$, we have that 
the probability $P^{l(n-1)}_{DT^{out}}$ that ``the specific token $T_{\lambda}^{out}$ is at node $v_i$ after $l(n-1)$ time steps'' is 
\begin{equation}\label{lowerProf_no_oscil_dyn}
P^{l(n-1)}_{DT^{out}} \geq (1+\mathcal{D}^+_{max})^{-l(n-1)} > 0 .
\end{equation}
This is mainly due to the fact that every $l$ time steps, each edge is active for at least one time step.  Since the nominal digraph $\mathcal{G}_d$ is strongly connected, it has a path of length at most $n-1$ from each node $v_l$ to each node $v_i$. 
Thus, at the first $l$ steps, we can select the first edge in this path (at the instant when it is active) and use self loops at the remaining instants; during the next $l$ time steps, we can select the second edge on this path and use self loops at the remaining instants; and so forth.

From \eqref{lowerProf_no_oscil_dyn} we have that the probability $P^{l(n-1)}_{N\_ DT^{out}}$ that ``the specific token $T_{\lambda}^{out}$ is not at node $v_i$ after $l(n-1)$ time steps'' is
\begin{equation}\label{lowerProf_not_no_oscil_dyn}
P^{l(n-1)}_{N\_ DT^{out}} \leq 1- (1+\mathcal{D}^+_{max})^{-l(n-1)} ,
\end{equation}
and from this point onward, we can use steps similar to the analysis in the proof of Theorem~\ref{Theorem2_prob_no_oscill}.   $ \hfill  \qed $
\end{pf}

\begin{remark}
Note that during the operation of Algorithm~\ref{algorithm_prob_no_oscil_dynamic} if we adopt the set of assumptions $A_1, \overline{A_2}$ for $\mathcal{G}_d[k]$ during every $k$, then \eqref{lowerProf_no_oscil_dyn} becomes 
\begin{equation}\label{Meet_prob_dyn_second}
P^{n-1}_{DT_i} \geq [ p_{\theta_{min}} (1+\mathcal{D}^+_{max}) ]^{-(n-1)} > 0,
\end{equation}
where $p_{\theta_{min}} = \min_{\theta \in \{1, 2, ..., M \}} p_{\theta}$. 
Since each digraph $G_{d_{\theta}}$, for $\theta \in \{1, 2, ..., M \}$, is selected in an i.i.d. manner with probability $p_\theta > p_{\theta_{min}}$ and the union graph is strongly connected, we can first select a topology that includes the first edge on the path from node $v_l$ to node $v_i$ (at least one such topology exists) then select a topology that includes the second edge on the path from node $v_l$ to node $v_i$ (at least one such topology exists), and so forth (with self loops included if necessary).
Then, \eqref{lowerProf_no_oscil_dyn} is replaced by \eqref{Meet_prob_dyn_second} and the structure of the proof remains the same. 
\end{remark}

It is important to note here that Algorithm~\ref{algorithm_prob_no_oscil_dynamic} converges in finite time even in the presence of dynamic communication topologies. 
Compared to Algorithm~\ref{algorithm_prob_no_oscil}, the main difference is an increase on the required number of time steps for convergence (which will be shown explicitly in the next section). 
However, in practical applications, there is also a possible increase in the processor usage of every node (due to the calculation of the nonzero probabilities $b_{lj}[k]$ for each of its outgoing edges $m_{lj}[k]$ during each time step $k$ in Iteration Step~$1$ of Algorithm~\ref{algorithm_prob_no_oscil_dynamic}) and a possible increase on the required number of transmissions for convergence. 
Analysis of the requirements on processor usage and number of transmissions will be considered in the future in order to highlight the proposed algorithms' operational advantages.

% ===============================================
%
%
% SIMULATIONS 
%
%
% ===============================================
\section{Simulation Results}\label{results}

In this section, we illustrate the behavior and the advantages of the proposed distributed algorithms. 
We execute Algorithm~\ref{algorithm_prob_no_oscil} over a static random digraph of $10$ nodes, and Algorithm~\ref{algorithm_prob_no_oscil_dynamic} over a set of dynamic digraphs of $10$ nodes whose union graph is equal to the nominal digraph after $l=5$ time steps. 
The initial quantized states of the nodes were randomly chosen between $1$ and $50$ (for each node, the initial state was a randomly chosen quantized value between $1$ and $50$ with probability $\dfrac{1}{50}$) with the average of the initial states of the nodes turning out to be $q = \dfrac{368}{10} = 36.8$ which means that $\lfloor q \rfloor = 36$ and $\lceil q \rceil = 37$.

\begin{figure*} [ht]
\centering
\includegraphics[width=55mm]{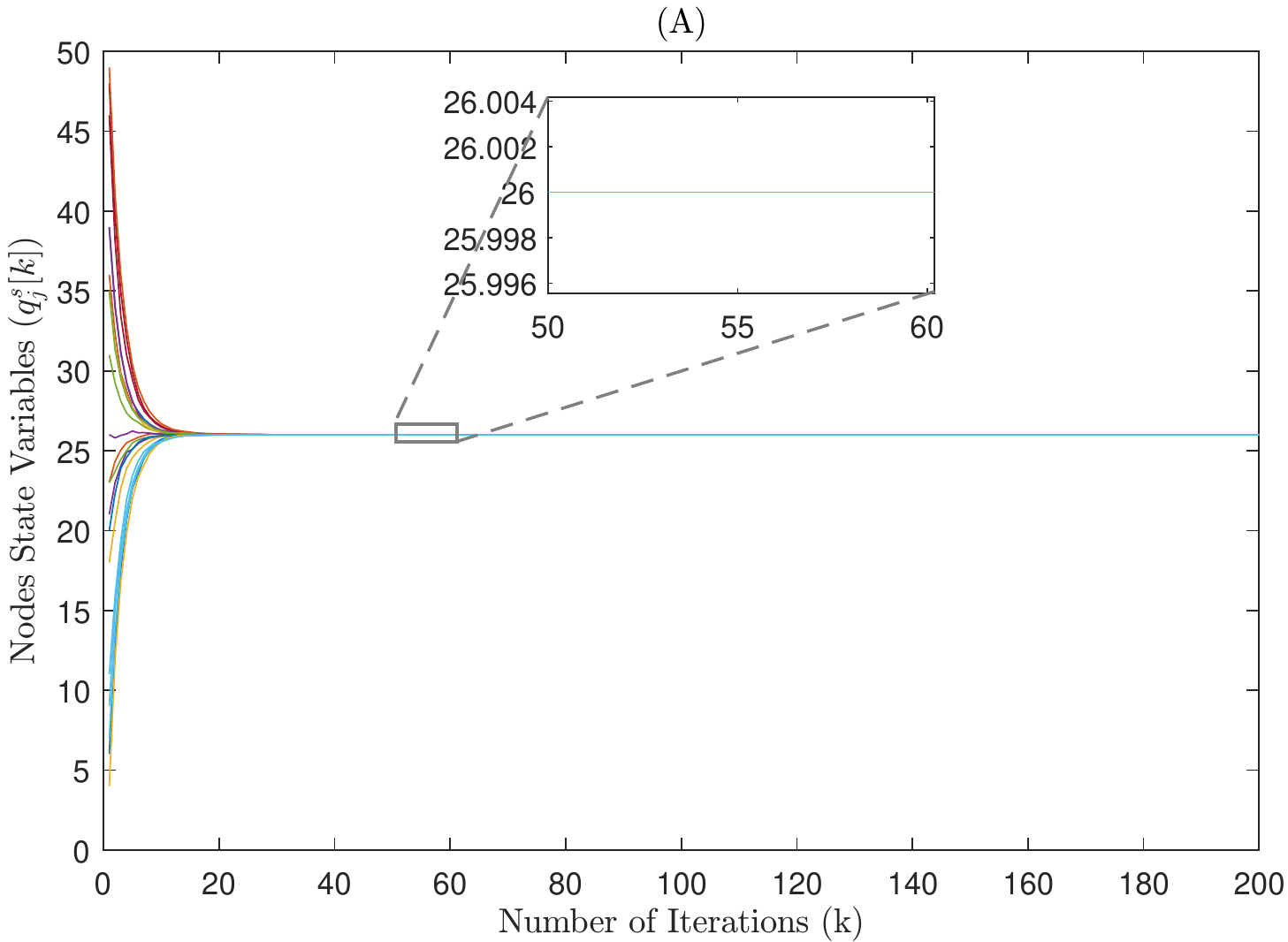}~~\hspace*{0.5cm}
\includegraphics[width=55mm]{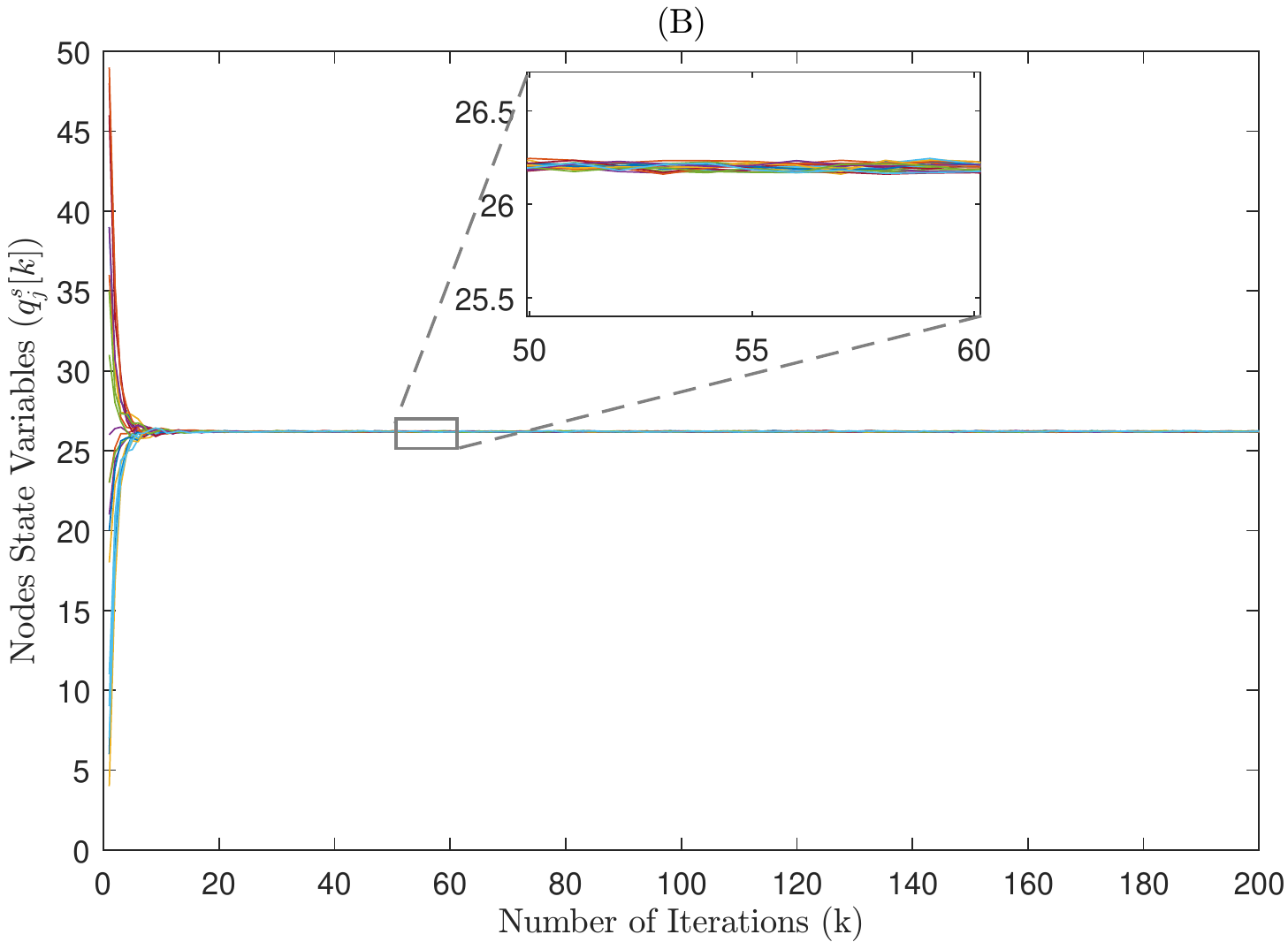}~~\hspace*{0.5cm}
\includegraphics[width=55mm]{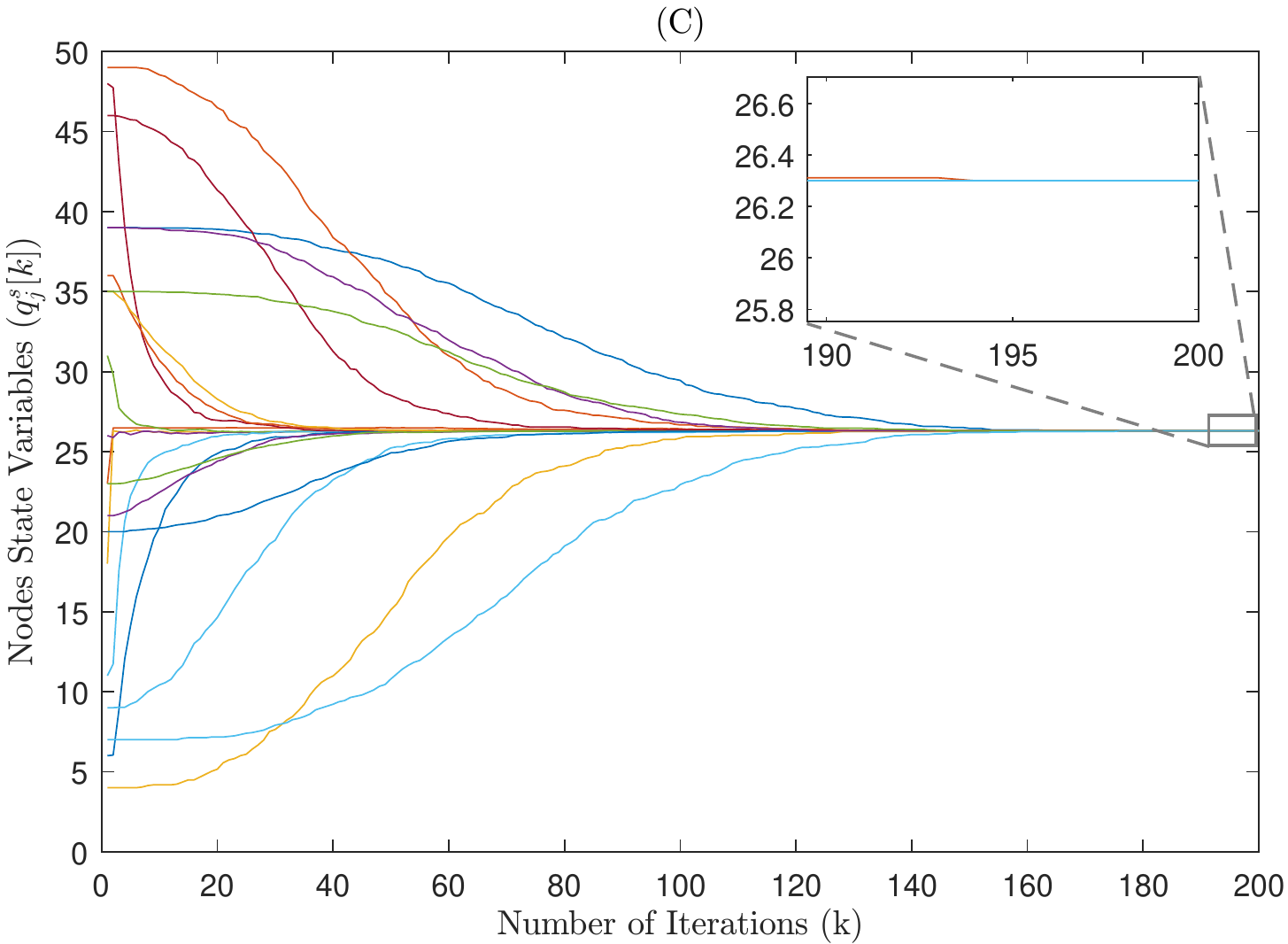} \\ \vspace{.2cm}
\includegraphics[width=55mm]{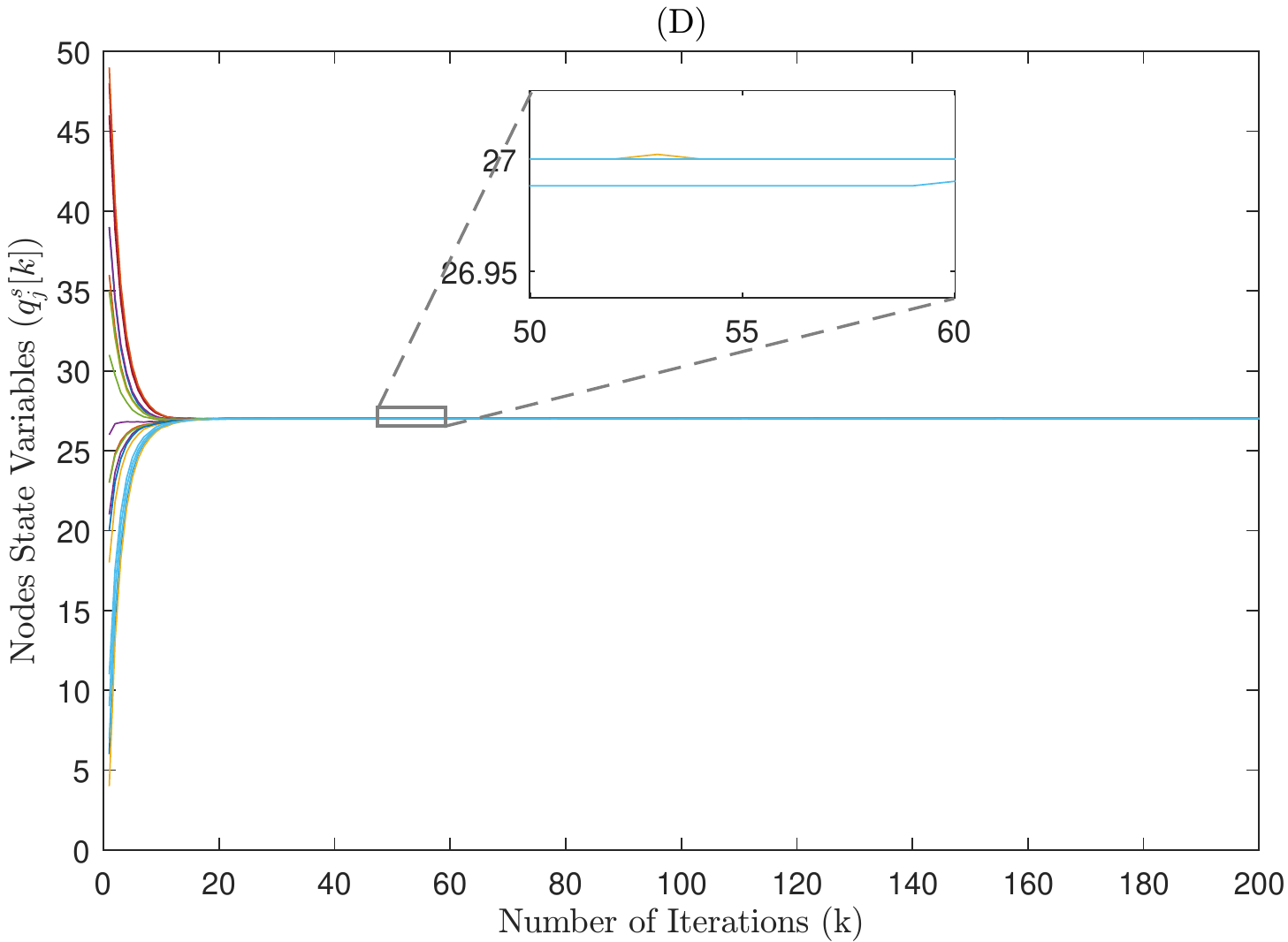}~~\hspace*{0.5cm}
\includegraphics[width=55mm]{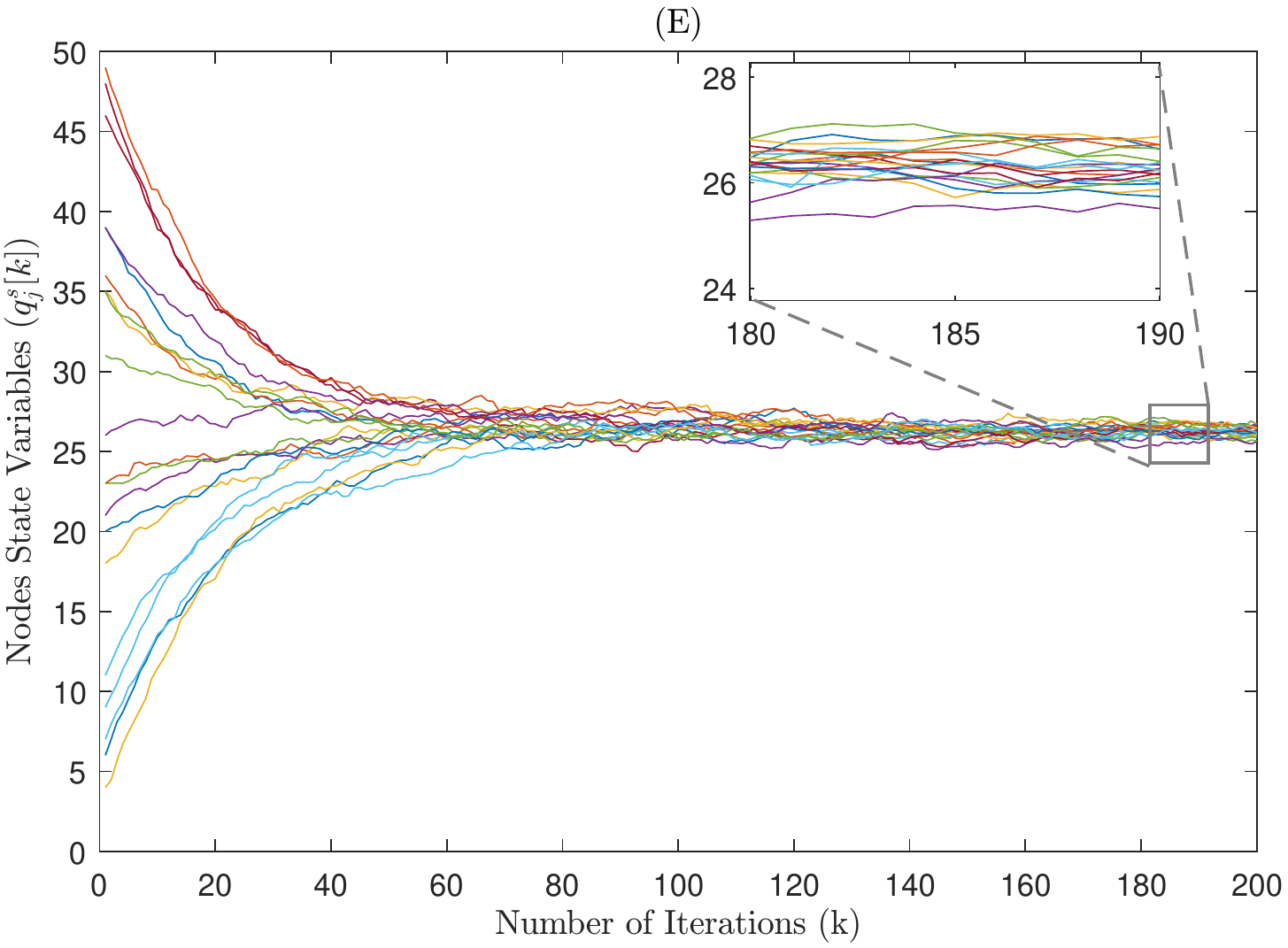}~~\hspace*{0.5cm}
\includegraphics[width=55mm]{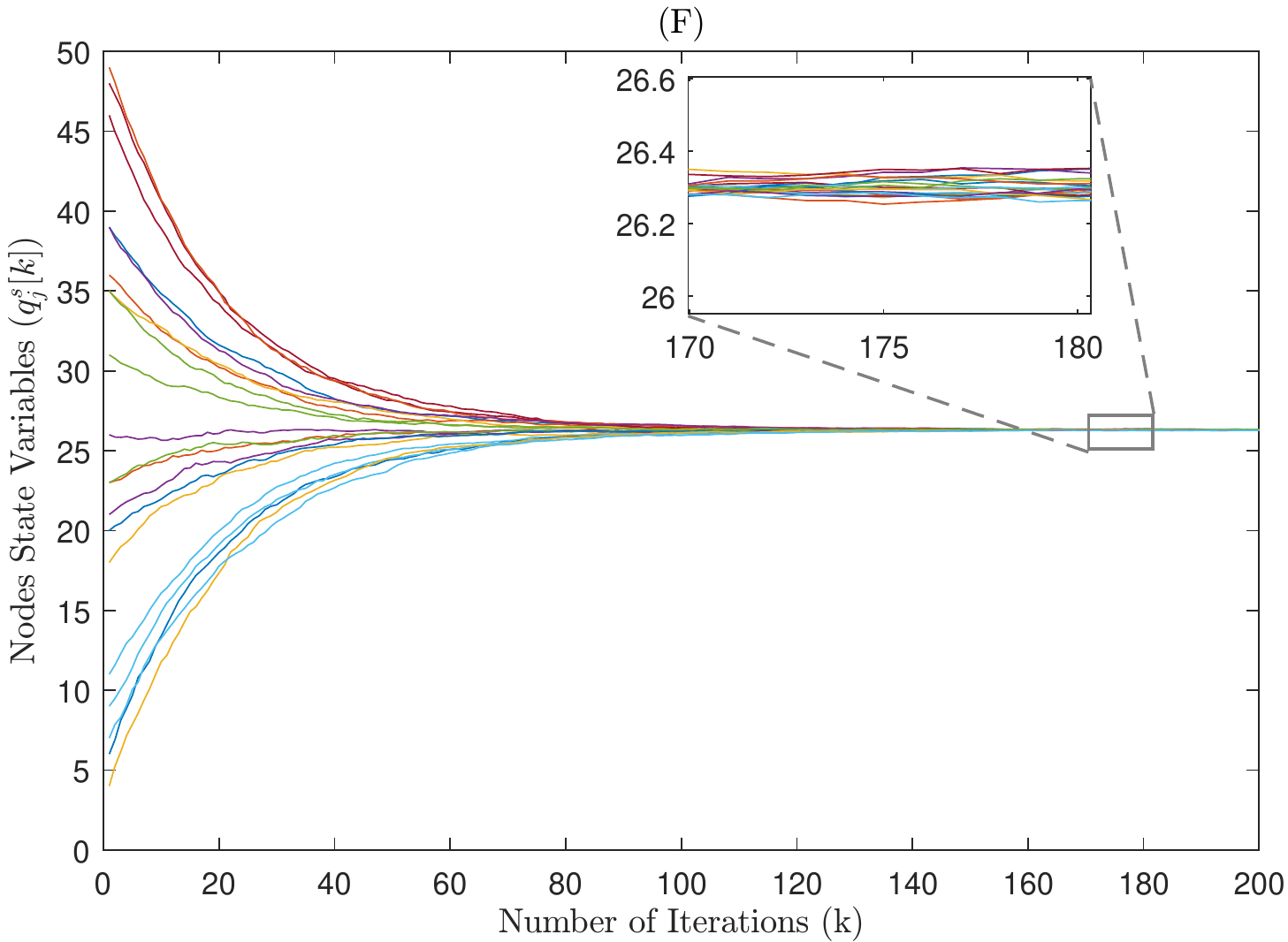}
\caption{Comparison between Algorithm~\ref{algorithm_prob_no_oscil} (A), the distributed averaging algorithm with quantized communication in \cite{2020:RikosHadj_IFAC} (B), the distributed averaging algorithm with quantized communication in \cite{2020:Rikos_Quant_Cons} (C), the distributed averaging algorithm with quantized communication in \cite{2016:Chamie_Basar} (D), the quantized asymmetric averaging algorithm in \cite{2011:Cai_Ishii} (E), and the quantized gossip algorithm in \cite{2007:Basar} (F), averaged over $1000$ randomly generated strongly connected digraphs of $20$ nodes each.}
\label{comp20}
\end{figure*}

Then, we show the average number of time steps needed for quantized average consensus to be reached over $1000$ randomly generated digraphs of $20$ nodes each and compare the performance of Algorithm~\ref{algorithm_prob_no_oscil} against existing state-of-the-art approaches. 
The initial quantized states of the nodes were also randomly chosen between $1$ and $50$ with the average of the initial states of the nodes turning out to be $q = \dfrac{526}{20} = 26.3$. 
Furthermore, for convenience, the initial quantized state of each node remained the same for each one of the $1000$ randomly generated digraphs, which means that the average of the nodes initial quantized states also remained equal to $q = \dfrac{526}{20} = 26.3$. 
We compare the performance of Algorithm~\ref{algorithm_prob_no_oscil} against five other algorithms: 
(a) the distributed averaging algorithm with quantized communication presented in \cite{2020:RikosHadj_IFAC} in which, at each time step $k$, each agent splits its mass variables in equal pieces and then transmits all of the pieces to randomly chosen out-neighbors, 
(b) the distributed averaging algorithm with quantized communication presented in \cite{2020:Rikos_Quant_Cons} in which, at each time step $k$, each agent sends its mass variables towards an out-neighbor chosen according to a priority in the form of a quantized fraction,
(c) the distributed averaging algorithm with quantized communication presented in \cite{2016:Chamie_Basar} in which, at each time step $k$, each agent $v_j$ broadcasts a quantized version of its own state towards its out-neighbors, 
(d) the quantized asymmetric averaging algorithm presented in \cite{2011:Cai_Ishii} in which, at each time step $k$, one edge, say edge $(v_l, v_j)$, is selected at random and, node $v_j$ sends its state information and surplus to node $v_l$, which performs updates over its own state and surplus values, 
(e) the quantized gossip algorithm presented in \cite{2007:Basar} in which, at each time step $k$, one edge\footnote{Note here that the algorithm in \cite{2007:Basar} requires the underlying graph to be undirected. For this reason, in Fig.~\ref{comp20}, for \cite{2007:Basar}, we make the randomly generated underlying digraphs undirected (by enforcing that if $(v_j, v_i) \in \mathcal{E}$ then also $(v_i, v_j) \in \mathcal{E}$) while, for the algorithms in \cite{2011:Cai_Ishii, 2016:Chamie_Basar, 2020:Rikos_Quant_Cons, 2020:RikosHadj_IFAC}, the randomly generated underlying graph is generally directed.} is selected at random, independently from earlier instants, and the states of the nodes that the selected edge is incident on are updated.

\begin{figure}[t]
\begin{center}
\includegraphics[width=0.90\columnwidth]{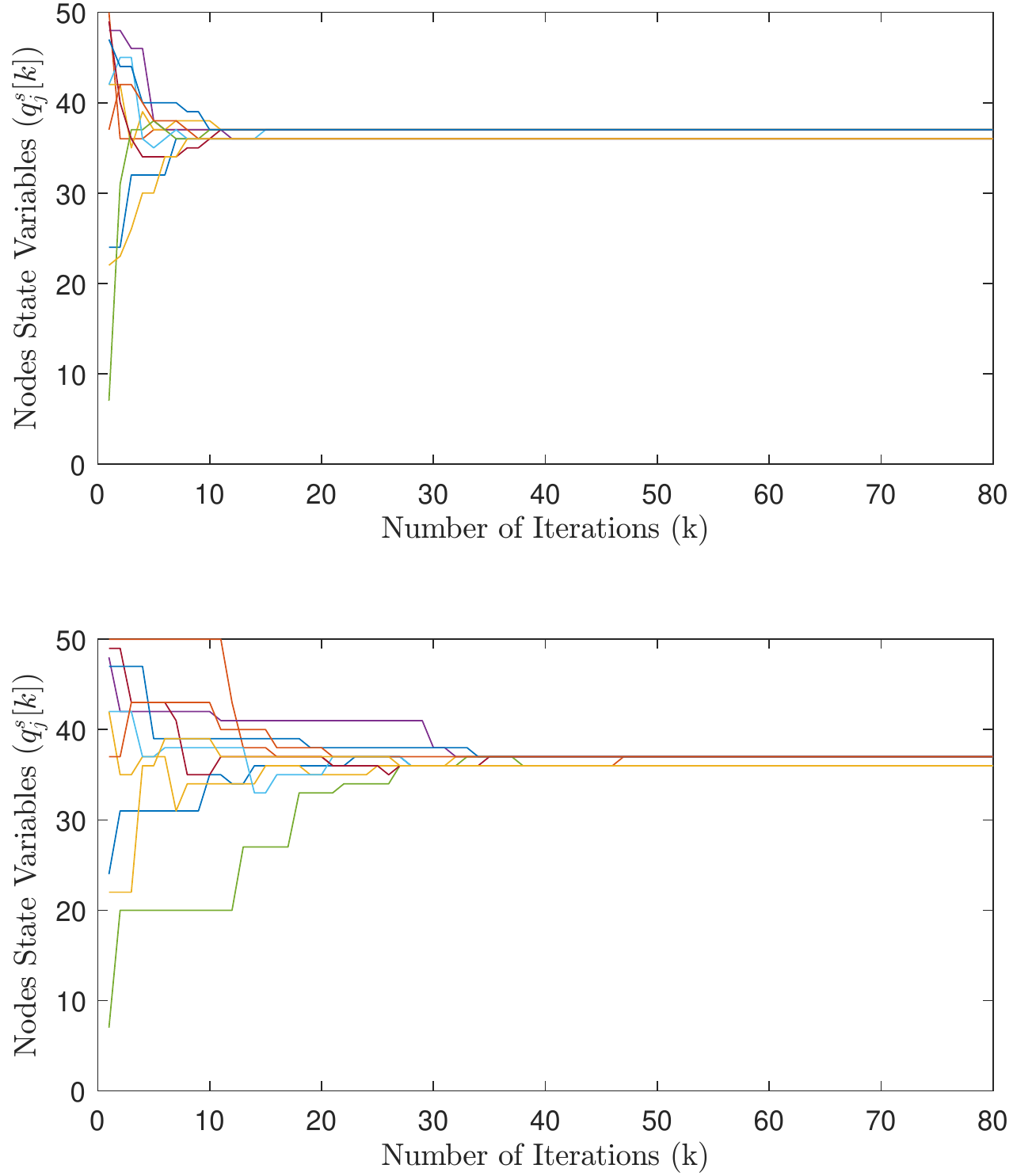}
\caption{Execution of Algorithm~\ref{algorithm_prob_no_oscil} over a Static Random Digraph of $10$ nodes (\textit{Top Figure}) and Algorithm~\ref{algorithm_prob_no_oscil_dynamic} over a Set of Dynamic Digraphs of $10$ nodes each (\textit{Bottom Figure}).}
\label{prob_example_figures}
\end{center}
\end{figure}

Fig.~\ref{prob_example_figures} shows the operation of Algorithm~\ref{algorithm_prob_no_oscil} over a static random digraph of $10$ nodes and Algorithm~\ref{algorithm_prob_no_oscil_dynamic} over a set of dynamic digraphs of $10$ nodes whose union is equal to the nominal digraph after $l=5$ time steps. 
In both cases the average of the initial states of the nodes being equal to $q = \dfrac{368}{10} = 36.8$. 
On the top of Fig.~\ref{prob_example_figures} we can see that during the operation of Algorithm~\ref{algorithm_prob_no_oscil} every node is able to reach quantized average consensus after $10$ time steps and the states of the nodes stabilize to be equal either to $\lfloor q \rfloor = 36$ or $\lceil q \rceil = 37$ after $15$ time steps. 
At the bottom of Fig.~\ref{prob_example_figures} we can see that Algorithm~\ref{algorithm_prob_no_oscil_dynamic} requires more steps to converge than Algorithm~\ref{algorithm_prob_no_oscil} due to the dynamic nature of the communication topology (since the union graph of the the set of dynamic digraphs is equal to the nominal digraph after $l=5$ time steps) and each node's state is able to stabilize to be equal either to $\lfloor q \rfloor = 36$ or $\lceil q \rceil = 37$ after $47$ time steps. 
This makes Algorithm~\ref{algorithm_prob_no_oscil_dynamic} the first algorithm in the literature to achieve oscillation-free quantized average consensus after a finite number of time steps over dynamic digraphs without any network requirements since (i) in \cite{2011:Cai_Ishii} the calculation of the quantized average relies on a static threshold that depends on the number of nodes in the network, (ii) in \cite{2016:Chamie_Basar} the operation requires a set of weights over the links of the dynamic digraph that form a doubly stochastic matrix which need to be recalculated again during each time step (see \cite{2014:RikosHadj, 2012:CortesJournal}) while the states of the nodes exhibit an oscillating behavior, and (iii) in \cite{2007:Basar} the operation requires bidirectional communication (i.e., undirected graph) and the states of the nodes also exhibit an oscillating behavior.

Fig.~\ref{comp20} shows the average number of time steps needed for quantized average consensus to be reached over $1000$ randomly generated digraphs of $20$ nodes each, in which the average of the nodes initial states is equal to $q = \dfrac{526}{20} = 26.3$. 
In Fig.~\ref{comp20} we can see that Algorithm~\ref{algorithm_prob_no_oscil} generally outperforms most finite time algorithms in the current literature. 
Its convergence speed is equal to \cite{2020:RikosHadj_IFAC} and \cite{2016:Chamie_Basar}, with the difference, however, being that every node's state is able to stabilize either to $\lfloor q \rfloor = 26$ or $\lceil q \rceil = 27$ rather than oscillate between these states. 
This can be seen from (A), (B), (D) in Fig.~\ref{comp20}. 
Specifically, during Algorithm~\ref{algorithm_prob_no_oscil} (see (A)) each node's state becomes equal to $\lfloor q \rfloor = 26$. 
However, during the algorithms in \cite{2020:RikosHadj_IFAC} and \cite{2016:Chamie_Basar} (see (B), (D)) the state of each node does not become equal to a specific value due to oscillations between $\lfloor q \rfloor = 26$ or $\lceil q \rceil = 27$.
The proposed algorithm has no prerequisites regarding the underlying communication topology (e.g., there is no need to obtain a set of weights over the digraph links that form a doubly stochastic matrix \cite{2016:Chamie_Basar}).

\section{Conclusions}\label{future}

We have considered the quantized average consensus problem and presented a randomized distributed averaging algorithm in which the processing, storing and exchange of information between neighboring agents is subject to uniform quantization. 
We analyzed its operation, established its correctness and showed that it allows every agent to reach a consensus state equal to either the ceiling or the floor of the real average (thus avoiding oscillating behavior) without any specific requirements regarding the network that describes the underlying communication topology, apart from strong connectedness (see \cite{2016:Chamie_Basar}).
Furthermore, we presented experimental results and argued that its convergence speed appears to outperform almost every algorithm in the available literature. 
Finally, we presented an enhanced version of our algorithm in which each agent achieves quantized average consensus (while also avoiding oscillating behavior) in the presence of a dynamically changing communication network.

In the future we plan to investigate stricter bounds on the convergence speed of Algorithms~\ref{algorithm_prob_no_oscil} and \ref{algorithm_prob_no_oscil_dynamic}. 
We also plan to explore their performance in the presence of network unreliability (e.g., delays and packet drops).

\vspace{-0.3cm}

% ------------------------------------------------------------------------------
% Bibliography
% ------------------------------------------------------------------------------

\bibliographystyle{plain}        % Include this if you use bibtex 
\bibliography{bibliografia_consensus}           % and a bib file to produce the 

\end{document}